\documentclass[journal]{IEEEtran}
\IEEEoverridecommandlockouts 

\makeatletter
\let\NAT@parse\undefined
\makeatother

\usepackage{amsmath,amssymb,amsfonts,amsthm,commath,esint}

\usepackage{bm} 
\usepackage[numbers,sort&compress]{natbib}

\usepackage{mathtools}
\usepackage{multirow}
\usepackage{physics}
\usepackage{color} 

\usepackage{graphicx}  
\usepackage{enumitem}  
\usepackage{optidef} 
\usepackage{mleftright}
\usepackage{cuted}
\usepackage{algorithm2e}
\usepackage{microtype}
\usepackage{ifthen} 
\usepackage{subcaption}
\usepackage{soul}
\usepackage{tikz}
\usepackage{graphicx} 
\usetikzlibrary{shapes.geometric, arrows.meta, positioning, calc, fit, backgrounds,shadings}
\usepackage{booktabs}
\usepackage{pgfplots}
\pgfplotsset{compat=1.18}
\usepgfplotslibrary{groupplots}

\usepackage[acronym]{glossaries}

\SetKwInput{kwInit}{Initialization}

\ifCLASSOPTIONcompsoc
    \usepackage[caption=false, font=normalsize, labelfont=sf, textfont=sf]{subfig}
\else
\usepackage[caption=false, font=footnotesize]{subfig}
\fi

\makeatletter
\def\blfootnote{\xdef\@thefnmark{}\@footnotetext}
\makeatother

\DeclarePairedDelimiterXPP\Aver[1]{\mathbb{E}}{[}{]}{}{

#1
}

\definecolor{purple}{rgb}{0.63, 0.13, 0.49}
\definecolor{sepia}{rgb}{0.44, 0.26, 0.08}
\definecolor{orange}{rgb}{1, 0.5, 0}
\definecolor{dkgreen}{rgb}{0.4, 0.2, 0.5}

\usepackage[normalem]{ulem}

\usepackage{pgfplots}
\usepackage[caption=false,font=footnotesize]{subfig}
\pgfplotsset{compat=1.18}

\begin{document}

\title{ Digital Twin-Guided Wireless Localization via  Probabilistic Inverse Learning
\thanks{
  Mahmoud Saad Abouamer and Henrik C. M. Frederiksen contributed equally to this work. \\
The authors are  with the Department of Electronic Systems, Aalborg University, Denmark. e-mail: $\{$mahmoudabo,hcmf,rjw,srp,petarp$\}$@es.aau.dk. \\
This work was supported in part by the Villum Investigator Grant ``WATER" from the Velux Foundation, Denmark. The work of  H.C.M. Frederiksen and S.R. 
Pandey was supported in part by DFF-Forsknings projekt “NETML” with grant No. 4286-00278B. 
    }
}
\author{
\IEEEauthorblockN{
Mahmoud Saad Abouamer,~\IEEEmembership{Member,~IEEE},
Henrik C. M. Frederiksen,
Robin J. Williams,~\IEEEmembership{Member,~IEEE},\\ Shashi Raj Pandey,~\IEEEmembership{Member,~IEEE}, Petar Popovski,~\IEEEmembership{Fellow,~IEEE} 
}
}

\maketitle

\begin{abstract}
This work investigates how site-specific digital twins (DTs) can support wireless localization using simplified distributed infrastructure. We propose \emph{PRISM} (Probabilistic Refinement using Inverse learning and a Site-specific Model), a DT-guided probabilistic localization scheme with single-antenna access points (APs). Localization is formulated as a probabilistic inverse problem in which a conditional proposal distribution over UE locations is learned from multi-node channel impulse responses (CIRs). The proposal generates candidate locations, with its spread regularized
during training and DT consistency used to reweight candidates and refine
the estimate at inference. The DT serves as an approximate forward-consistency model, avoiding assumptions of exact channel fidelity or differentiability with respect to user position. Numerical experiments demonstrate gains over data-driven and classical fingerprinting benchmarks, while achieving a median error of approximately $0.7$ m using 10 single-antenna APs. Without retraining, PRISM remains effective under degraded inference-time DT features. PRISM also yields uncertainty estimates from its proposal covariance, with empirical $\chi^2$ calibration approaching ideal behavior as APs increase.

\end{abstract}


\section{Introduction}

The increasing computing capabilities of the wireless infrastructure as well as the requirements of 6G applications drive the emergence of accurate site-specific wireless localization \cite{positioning_iot,pos_sustainability,5G_localization_tutorial,THz_localization_tut}. However, inferring location from wireless channel observations remains challenging due to noise, complex multipath propagation, and strict requirements, including low hardware complexity and cost efficiency \cite{positioning_iot,loc_challenges_2,loc_challenges}.

Conventional approaches often address these challenges by improving channel-observation resolution through more complex sensing hardware. For instance, massive multiple-input multiple-output (MIMO) systems and large antenna arrays provide finer spatial resolution, enabling multipath components to be more effectively separated and identified \cite{localization_massive_MIMO,localization_massive_MIMO_2,Channel_radio_env}. This reduces location ambiguity by making channel responses more distinguishable across candidate positions. However, these gains come at the cost of increased hardware complexity, energy consumption, and deployment overhead, which can limit scalability in practical industrial environments~\cite{D_MIMO_localization_impt}.

Instead of relying on complex sensing setups, this work investigates how site-specific propagation structure can support localization with simplified hardware. In particular, we consider a conservative distributed sensing setting where channel information is collected from a limited number of single-antenna access points, rather than from antenna arrays or dense fingerprint infrastructure. This setting is not meant to argue against richer sensing hardware, but to isolate the value of DT-guided probabilistic refinement when the channel observations are limited. The central challenge is then to exploit site-specific physical knowledge from these limited observations while avoiding restrictive assumptions on the digital twin, such as exact channel accuracy or differentiability with respect to user position. To address this challenge, we leverage a site-specific digital twin (DT) \cite{DT_general,DT_AK_general,khan2022digital,AK_DT_vision} of the wireless channel as an approximate forward-consistency check that scores candidate locations based on the propagation structure of the deployment environment.

By mapping channel observations or derived features directly to spatial coordinates, learning-based localization models are commonly used to capture environment-dependent structure from the data without requiring an explicit propagation model \cite{survey_localization_complexity,CNN_fingerprinting_like_transformer_extension_heavy_on_OG,CNN_fingerprinting_like_OG,CNN_fingerprinting_OG,another_transformer_localization,direct_localization_AE}. Early works based on convolutional neural networks (CNNs) \cite{CNN_fingerprinting_OG} demonstrated the potential of such approaches, while more recent studies have adopted expressive architectures such as Transformers to learn richer channel representations \cite{CNN_fingerprinting_like_transformer_extension_heavy_on_OG,another_transformer_localization,CNN_fingerprinting_like_OG}. However, these methods typically formulate localization as a direct observation-to-position mapping and provide no explicit mechanism for enforcing consistency with the site-specific propagation physics. Consequently, predictions may be statistically plausible yet inconsistent with the site-specific propagation geometry.

\def\Building#1#2#3#4#5#6#7{%
    \path[fill=#7!35, draw=#7!70!black, thick]
        (#1,#2,#3) -- ++(#4,0,0) -- ++(0,0,#6) -- ++(-#4,0,0) -- cycle;
    \path[fill=#7!24, draw=#7!70!black, thick]
        (#1,#2,#3) ++(#4,0,0) -- ++(0,#5,0) -- ++(0,0,#6) -- ++(0,-#5,0) -- cycle;
    \path[fill=#7!13, draw=#7!70!black, thick]
        (#1,#2,#3) ++(0,0,#6) -- ++(#4,0,0) -- ++(0,#5,0) -- ++(-#4,0,0) -- cycle;

    \foreach \wx in {0.18,0.48,0.78} {
        \foreach \wz in {0.25,0.55,0.80} {
            \path[fill=cyan!18, draw=#7!60!black, line width=0.25pt]
                (#1+\wx*#4,#2,#3+\wz*#6)
                -- ++(0.13*#4,0,0)
                -- ++(0,0,0.12*#6)
                -- ++(-0.13*#4,0,0)
                -- cycle;
        }
    }

    \foreach \wy in {0.22,0.55,0.78} {
        \foreach \wz in {0.28,0.58,0.82} {
            \path[fill=cyan!12, draw=#7!60!black, line width=0.25pt]
                (#1+#4,#2+\wy*#5,#3+\wz*#6)
                -- ++(0,0.12*#5,0)
                -- ++(0,0,0.11*#6)
                -- ++(0,-0.12*#5,0)
                -- cycle;
        }
    }
}

\def\DTBuilding#1#2#3#4#5#6#7{%
    \path[fill=#7!22, draw=#7!70!black, thick]
        (#1,#2,#3) -- ++(#4,0,0) -- ++(0,0,#6) -- ++(-#4,0,0) -- cycle;
    \path[fill=#7!14, draw=#7!70!black, thick]
        (#1,#2,#3) ++(#4,0,0) -- ++(0,#5,0) -- ++(0,0,#6) -- ++(0,-#5,0) -- cycle;
    \path[fill=#7!8, draw=#7!70!black, thick]
        (#1,#2,#3) ++(0,0,#6) -- ++(#4,0,0) -- ++(0,#5,0) -- ++(-#4,0,0) -- cycle;
}

\def\Car#1#2#3#4{%
    \path[fill=#4!45, draw=#4!70!black, thick]
        (#1,#2,#3) -- ++(0.62,0,0) -- ++(0,0,0.18) -- ++(-0.62,0,0) -- cycle;
    \path[fill=#4!30, draw=#4!70!black, thick]
        (#1,#2,#3) ++(0.62,0,0) -- ++(0,0.32,0) -- ++(0,0,0.18) -- ++(0,-0.32,0) -- cycle;
    \path[fill=#4!22, draw=#4!70!black, thick]
        (#1,#2,#3) ++(0,0,0.18) -- ++(0.62,0,0) -- ++(0,0.32,0) -- ++(-0.62,0,0) -- cycle;

    \path[fill=#4!30, draw=#4!70!black, thick]
        (#1+0.16,#2,#3+0.18) -- ++(0.30,0,0) -- ++(0,0,0.14) -- ++(-0.30,0,0) -- cycle;
    \path[fill=#4!20, draw=#4!70!black, thick]
        (#1+0.46,#2,#3+0.18) -- ++(0,0.22,0) -- ++(0,0,0.14) -- ++(0,-0.22,0) -- cycle;

    \path[fill=cyan!18, draw=#4!60!black, line width=0.25pt]
        (#1+0.21,#2,#3+0.22) -- ++(0.09,0,0) -- ++(0,0,0.07) -- ++(-0.09,0,0) -- cycle;
    \path[fill=cyan!18, draw=#4!60!black, line width=0.25pt]
        (#1+0.34,#2,#3+0.22) -- ++(0.09,0,0) -- ++(0,0,0.07) -- ++(-0.09,0,0) -- cycle;

    \fill[black!80] (#1+0.13,#2,#3-0.02) circle (0.045);
    \fill[black!80] (#1+0.50,#2,#3-0.02) circle (0.045);
}

\begin{figure*}[t!]
\centering
\resizebox{0.99\textwidth}{!}{%
\begin{tikzpicture}[
    x=1cm,y=1cm,
    font=\sffamily\small,
    >=Stealth,
    panel/.style={draw=gray!55, rounded corners=7pt, thick, fill=gray!2},
    subpanel/.style={draw=gray!45, rounded corners=5pt, thick, fill=white},
    dtsubpanel/.style={draw=teal!55!black, rounded corners=5pt, thick, fill=teal!3},
    title/.style={font=\bfseries\normalsize, align=center},
    note/.style={font=\scriptsize, align=center},
    smallnote/.style={font=\footnotesize, align=center},
    ap/.style={circle, draw=black!80, fill=white, thick, inner sep=1.2pt},
    ue/.style={circle, draw=black!80, fill=orange!75, thick, inner sep=2.0pt},
    truth/.style={circle, draw=orange!85!black, fill=orange!60, very thick, inner sep=2.5pt},
    cand/.style={circle, draw=teal!75!black, fill=white, thick, inner sep=2.0pt},
    est/.style={circle, draw=red!80!black, fill=red!55, very thick, inner sep=2.3pt},
    ray/.style={dashed, thick, draw=blue!70!black},
    reflray/.style={dotted, thick, draw=teal!70!black},
    dtray/.style={dashed, thick, draw=orange!85!black},
    weakdtray/.style={dotted, thick, draw=teal!70!black}
]

\draw[panel] (0,0.15) rectangle (5.35,7.35);
\node[title] at (2.675,7.10) {(a) Physical Scene};

\draw[subpanel] (0.20,0.34) rectangle (5.15,6.88);

\begin{scope}
\clip (0.28,0.42) rectangle (5.08,6.80);
\begin{scope}[shift={(0.00,2.96)}, x={(0.70cm,-0.31cm)}, y={(0.70cm,0.31cm)}, z={(0cm,0.70cm)}]

\path[fill=green!8, draw=gray!55, thick] (0,0,0) -- (6.6,0,0) -- (6.6,4.6,0) -- (0,4.6,0) -- cycle;
\path[fill=gray!25, opacity=0.95] (0.10,2.00,0.01) -- (6.45,2.00,0.01) -- (6.45,2.58,0.01) -- (0.10,2.58,0.01) -- cycle;
\path[fill=gray!25, opacity=0.95] (3.00,0.10,0.012) -- (3.58,0.10,0.012) -- (3.58,4.45,0.012) -- (3.00,4.45,0.012) -- cycle;

\Building{0.50}{3.15}{0}{1.15}{0.82}{0.95}{orange}
\Building{4.35}{3.05}{0}{1.30}{0.85}{1.05}{blue}
\Building{1.20}{0.60}{0}{1.10}{0.85}{0.88}{teal}

\Car{3.80}{2.08}{0.02}{orange}

\draw[black!70, thick] (0.25,4.20,0) -- (0.25,4.20,1.25);
\node[ap] (pap1) at (0.25,4.20,1.25) {};
\node[note, fill=white, inner sep=1pt] at (0.25,4.20,1.55) {AP1};

\draw[black!70, thick] (6.35,4.20,0) -- (6.35,4.20,1.25);
\node[ap] (pap2) at (6.35,4.20,1.25) {};

\draw[black!70, thick] (0.25,0.25,0) -- (0.25,0.25,1.10);
\node[ap] (pap3) at (0.25,0.25,1.10) {};
\node[note, fill=white, inner sep=1pt] at (0.3,0.4,1.3) {AP3};

\draw[black!70, thick] (6.35,0.25,0) -- (6.35,0.25,1.10);
\node[ap] (pap4) at (6.35,0.25,1.10) {};
\node[note, fill=white, inner sep=1pt] at (6.35,0.25,0.78) {AP4};

\node[ue] (pue) at (3.35,2.28,0.22) {};
\node[note, fill=white, inner sep=1.1pt, rounded corners=2pt]
    at (4.18,1.95,0.44) {UE at $\boldsymbol u$};

\draw[ray] (pue) -- (pap1);
\draw[ray] (pue) -- (pap2);
\draw[ray] (pue) -- (pap3);
\draw[ray] (pue) -- (pap4);
\draw[reflray] (pue) -- (1.70,3.62,0.60) -- (pap1);
\draw[reflray] (pue) -- (5.10,3.48,0.68) -- (pap2);

\end{scope}
\end{scope}

\draw[panel] (5.55,0.15) rectangle (11.65,7.35);
\node[title] at (8.60,7.10) {(b) Belief over candidate locations};

\draw[dtsubpanel] (5.77,0.34) rectangle (11.43,6.88);

\begin{scope}
\clip (5.85,0.42) rectangle (11.35,6.80);
\begin{scope}[shift={(5.15,2.82)}, x={(0.78cm,-0.34cm)}, y={(0.78cm,0.34cm)}, z={(0cm,0.78cm)}]

\path[fill=teal!5, draw=teal!50!black, thick] (0,0,0) -- (6.6,0,0) -- (6.6,4.6,0) -- (0,4.6,0) -- cycle;
\path[fill=gray!20, opacity=0.95] (0.10,2.00,0.01) -- (6.45,2.00,0.01) -- (6.45,2.58,0.01) -- (0.10,2.58,0.01) -- cycle;
\path[fill=gray!20, opacity=0.95] (3.00,0.10,0.012) -- (3.58,0.10,0.012) -- (3.58,4.45,0.012) -- (3.00,4.45,0.012) -- cycle;

\DTBuilding{0.50}{3.15}{0}{1.15}{0.82}{0.95}{teal}
\DTBuilding{4.35}{3.05}{0}{1.30}{0.85}{1.05}{teal}
\DTBuilding{1.20}{0.60}{0}{1.10}{0.85}{0.88}{teal}
\DTBuilding{3.80}{2.08}{0.02}{0.62}{0.32}{0.18}{teal}

\draw[black!70, thick] (0.25,4.20,0) -- (0.25,4.20,1.25);
\node[ap] at (0.25,4.20,1.25) {};
\draw[black!70, thick] (6.35,4.20,0) -- (6.35,4.20,1.25);
\node[ap] at (6.35,4.20,1.25) {};
\draw[black!70, thick] (0.25,0.25,0) -- (0.25,0.25,1.10);
\node[ap] at (0.25,0.25,1.10) {};
\draw[black!70, thick] (6.35,0.25,0) -- (6.35,0.25,1.10);
\node[ap] at (6.35,0.25,1.10) {};

\draw[blue!60!black, very thick, dashed] (3.35,2.28,0.05) ellipse [x radius=1.15, y radius=0.58];
\draw[blue!45!black, thick, dashed, opacity=0.55] (3.35,2.28,0.05) ellipse [x radius=1.70, y radius=0.92];

\node[truth] (btrue) at (3.35,2.28,0.22) {};
\node[cand]  (bv1) at (3.92,1.95,0.22) {};
\node[cand]  (bv2) at (2.48,1.96,0.22) {};
\node[cand]  (bv3) at (4.18,2.72,0.22) {};
\node[cand]  (bv4) at (2.95,2.92,0.22) {};
\node[cand]  (bv5) at (3.68,2.98,0.22) {};
\node[cand]  (bv6) at (4.55,2.22,0.22) {};

\node[note, fill=white, inner sep=1.1pt, rounded corners=2pt, text=blue!70!black]
    at (1.45,3.40,0.48) {uncertain\\belief region};

\node[note, fill=white, inner sep=1.0pt, rounded corners=2pt, text=orange!85!black] (lblbtrue)
    at (2.45,1.28,0.42) {true $\boldsymbol u$};
\draw[->, thin, draw=orange!85!black] (lblbtrue.north east) -- (btrue);

\node[note, fill=white, inner sep=1.0pt, rounded corners=2pt] (lblbvone)
    at (4.92,1.34,0.42) {$\boldsymbol v_1$};
\draw[->, thin, draw=teal!70!black] (lblbvone.west) -- (bv1);

\node[note, fill=white, inner sep=1.0pt, rounded corners=2pt] (lblbvtwo)
    at (1.72,1.38,0.42) {$\boldsymbol v_2$};
\draw[->, thin, draw=teal!70!black] (lblbvtwo.east) -- (bv2);

\node[note, fill=white, inner sep=1.0pt, rounded corners=2pt] (lblbvthree)
    at (5.10,3.30,0.42) {$\boldsymbol v_3$};
\draw[->, thin, draw=teal!70!black] (lblbvthree.west) -- (bv3);

\end{scope}
\end{scope}

\draw[panel] (11.85,0.15) rectangle (20.10,7.35);
\node[title] at (15.975,7.10) {(c) DT forward check and refinement};

\draw[dtsubpanel] (12.07,0.34) rectangle (19.83,6.88);

\begin{scope}
\clip (12.15,0.42) rectangle (19.75,6.80);
\begin{scope}[shift={(11.20,2.82)}, x={(0.96cm,-0.42cm)}, y={(0.96cm,0.42cm)}, z={(0cm,0.96cm)}]

\path[fill=teal!5, draw=teal!50!black, thick] (0,0,0) -- (6.6,0,0) -- (6.6,4.6,0) -- (0,4.6,0) -- cycle;
\path[fill=gray!20, opacity=0.95] (0.10,2.00,0.01) -- (6.45,2.00,0.01) -- (6.45,2.58,0.01) -- (0.10,2.58,0.01) -- cycle;
\path[fill=gray!20, opacity=0.95] (3.00,0.10,0.012) -- (3.58,0.10,0.012) -- (3.58,4.45,0.012) -- (3.00,4.45,0.012) -- cycle;

\DTBuilding{0.50}{3.15}{0}{1.15}{0.82}{0.95}{teal}
\DTBuilding{4.35}{3.05}{0}{1.30}{0.85}{1.05}{teal}
\DTBuilding{1.20}{0.60}{0}{1.10}{0.85}{0.88}{teal}
\DTBuilding{3.80}{2.08}{0.02}{0.62}{0.32}{0.18}{teal}

\draw[black!70, thick] (0.25,4.20,0) -- (0.25,4.20,1.25);
\node[ap] (cap1) at (0.25,4.20,1.25) {};
\draw[black!70, thick] (6.35,4.20,0) -- (6.35,4.20,1.25);
\node[ap] (cap2) at (6.35,4.20,1.25) {};
\draw[black!70, thick] (0.25,0.25,0) -- (0.25,0.25,1.10);
\node[ap] (cap3) at (0.25,0.25,1.10) {};
\draw[black!70, thick] (6.35,0.25,0) -- (6.35,0.25,1.10);
\node[ap] (cap4) at (6.35,0.25,1.10) {};

\draw[blue!45!black, thick, dashed, opacity=0.45] (3.35,2.28,0.05) ellipse [x radius=1.70, y radius=0.92];

\node[truth] (ctrue) at (3.35,2.28,0.22) {};
\node[est]   (uhat)  at (3.00,2.18,0.22) {};

\node[cand] (cv1) at (3.92,1.95,0.22) {};
\node[cand] (cv2) at (2.48,1.96,0.22) {};
\node[cand] (cv3) at (4.18,2.72,0.22) {};
\node[cand] (cv4) at (2.95,2.92,0.22) {};
\node[cand] (cv5) at (3.68,2.98,0.22) {};
\node[cand] (cv6) at (4.55,2.22,0.22) {};

\draw[dtray] (cv1) -- (cap1);
\draw[dtray] (cv1) -- (cap2);
\draw[dtray] (cv1) -- (cap3);
\draw[dtray] (cv1) -- (cap4);

\draw[weakdtray] (cv2) -- (cap1);
\draw[weakdtray] (cv2) -- (cap3);
\draw[weakdtray] (cv3) -- (cap2);
\draw[weakdtray] (cv3) -- (cap4);
\draw[weakdtray] (cv4) -- (cap1);
\draw[weakdtray] (cv6) -- (cap4);

\node[note, fill=white, inner sep=1.0pt, rounded corners=2pt, text=orange!85!black] (lbltrue)
    at (1.65,2.18,0.48) {true $\boldsymbol u$};
\draw[->, thin, draw=orange!85!black] (lbltrue.east) -- (ctrue);

\node[note, fill=white, inner sep=1.0pt, rounded corners=2pt] (lblvtwo)
    at (1.18,2.98,0.48) {$\boldsymbol v_2$: low $w_2$};
\draw[->, thin, draw=teal!70!black] (lblvtwo.east) -- (cv2);

\node[note, fill=white, inner sep=1.0pt, rounded corners=2pt] (lblvone)
    at (4.75,1.32,0.50) {$\boldsymbol v_1$: high $w_1$};
\draw[->, thin, draw=orange!85!black] (lblvone.west) -- (cv1);

\node[note, fill=white, inner sep=1.0pt, rounded corners=2pt] (lblvthree)
    at (5.00,3.22,0.52) {$\boldsymbol v_3$: low $w_3$};
\draw[->, thin, draw=teal!70!black] (lblvthree.west) -- (cv3);

\node[note, fill=white, inner sep=1.0pt, rounded corners=2pt, text=red!75!black] (lbluhat)
    at (1.20,1.02,0.46) {refined $\hat{\boldsymbol u}$};
\draw[->, thin, draw=red!75!black] (lbluhat.east) -- (uhat);

\end{scope}
\end{scope}

\end{tikzpicture}%
}
\caption{Illustration of DT-guided probabilistic localization. 
(a) A UE at unknown location $\boldsymbol u$ transmits an uplink localization signal, and distributed single-antenna APs observe the resulting multipath channel feature $\boldsymbol y_{\mathrm{obs}}$. 
(b) Localization is represented as an uncertain belief over plausible UE locations in the site-specific DT. For illustration, the true location is shown in orange, and the white points denote candidate locations $\{\boldsymbol v_i\}$ sampled from the belief. 
(c) The DT is used in the forward, location-to-channel direction: each candidate location $\boldsymbol v_i$ is queried to produce a predicted channel feature $\hat{\mathcal G}(\boldsymbol v_i)$, which is then checked for consistency with $\boldsymbol y_{\mathrm{obs}}$. Candidate $\boldsymbol v_1$, which lies close to the true location, receives the largest weight, while $\boldsymbol v_2$ and $\boldsymbol v_3$ receive smaller weights. The red marker denotes the refined estimate $\hat{\boldsymbol u}$. In PRISM, candidate sampling is regularized during training, while the DT forward check is used to reweight candidates and refine the location estimate during inference.}

\label{fig:prism_intro_concept}
\end{figure*}
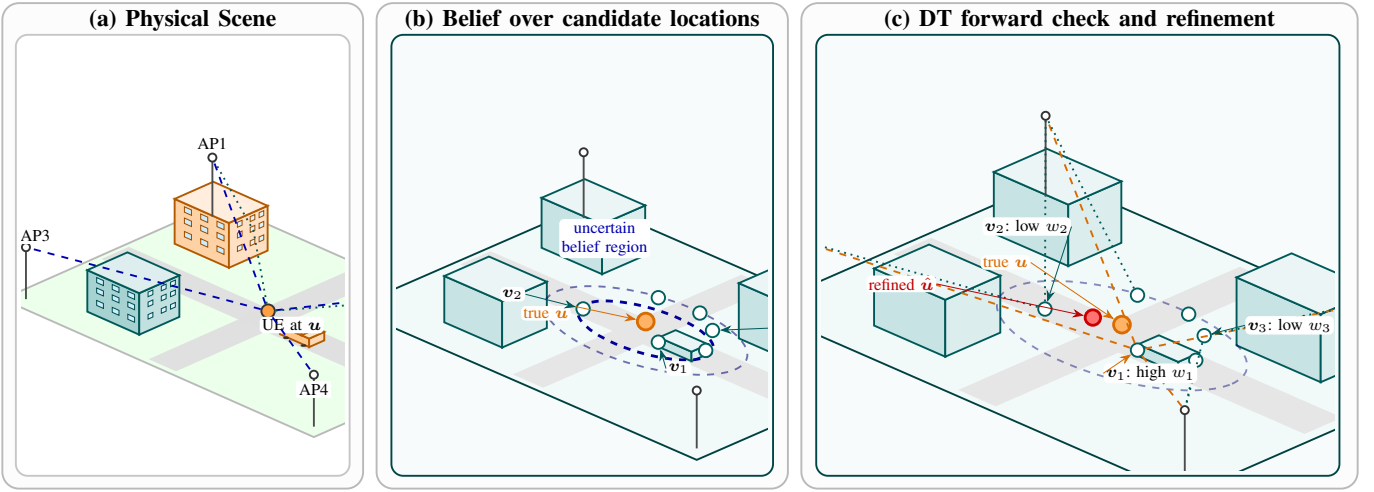

In contrast to the direct location-estimation approaches described above, soft-information-based localization constructs probabilistic information from standardized time- and angle-based 5G measurements, such as downlink time-difference-of-arrival measurements, rather than relying only on single-value estimates \cite{Self_information_localization}. While effective in cellular localization settings, this approach relies on learned statistical measurement models and does not explicitly exploit site-specific propagation geometry through a DT. Deep variational learning has also been applied to multipath-assisted single-anchor localization, where ultra-wideband channel impulse responses are used to estimate position-related parameters, including distance, time-difference-of-arrival, and angle- or phase-related quantities, before geometric localization \cite{Deepvariational_localization}. This improves parameter estimation under non-line-of-sight conditions, but targets an array-equipped single-anchor setting rather than distributed single-antenna sensing and does not use site-specific DT feedback to guide localization.

To incorporate site-specific physical knowledge, recent works have explored DTs and ray-tracing tools for wireless localization \cite{DT_localization_AK,DT_RT_phase}. By capturing environment-specific structures such as blockages, reflections, and dominant propagation paths, DTs provide a principled way to incorporate physical knowledge into localization. Existing approaches, however, primarily use DTs to generate synthetic fingerprinting datasets, augment training data, or learn neural surrogates of propagation \cite{DT_localization_AK,Neural_Augmentted_Localization}. In such cases, the DT mainly supports data generation or model training, rather than directly constraining the location estimate through channel-domain consistency during inference. Recent work in \cite{han2026rayloc} has used dedicated differentiable ray-tracing formulations to directly optimize point estimates of user position via iterative gradient-based methods. Such approaches rely on useful derivatives of the channel response with respect to spatial coordinates, which may require simplified propagation models, while the iterative position optimization can increase computational overhead and limit applicability in realistic environments.

Taken together, existing approaches either rely on learned neural mappings without explicit propagation constraints or use DTs mainly for data generation and model training. This leaves open the problem of using site-specific physical structure to guide localization without requiring restrictive assumptions such as exact channel accuracy or differentiability with respect to user position, while preserving compatibility with efficient ray-tracing tools such as Sionna~\cite{Sionna_Intro_Journal,Sionna_SF_scene}. 
Fig.~\ref{fig:prism_intro_concept} illustrates the principle pursued in this work. A UE transmits an uplink localization signal, while a limited number of distributed single-antenna APs passively observe the resulting multi-AP CIR feature $\boldsymbol y_{\mathrm{obs}}$. The localization task is inverse, since it seeks to infer the unknown UE position $\boldsymbol u$ from $\boldsymbol y_{\mathrm{obs}}$. By contrast, the DT is used in the forward, location-to-channel direction: given a hypothesized candidate location $\boldsymbol v_i$, the DT predicts the corresponding channel feature $\hat{\mathcal G}(\boldsymbol v_i)$, which can then be checked for consistency with $\boldsymbol y_{\mathrm{obs}}$. Thus, the DT is used only as an approximate forward-consistency model, which relaxes the need for differentiability with respect to UE position or for solving the inverse channel-to-location problem directly.

Building on this view, we propose \emph{PRISM}
(\textbf{P}robabilistic \textbf{R}efinement using
\textbf{I}nverse learning and a \textbf{S}ite-specific
\textbf{M}odel), a DT-guided probabilistic localization
framework for distributed sensing. The main
contributions are summarized as follows:
\begin{itemize}[leftmargin=*]
    \item We formulate localization as a probabilistic inverse
    problem in which an inverse proposal generates plausible UE
    locations and a site-specific DT evaluates their channel-domain
    consistency. Since the DT is queried only in the forward,
    location-to-channel direction, the framework does not require
    an exact or differentiable DT.

\item We develop a finite-candidate learning and inference procedure in which the learned proposal defines a data-informed candidate region. During
training, proposal regularization controls the candidate-search spread,
while during inference DT consistency weights explicitly reweight sampled
candidates.

    \item Through extensive numerical experiments, we demonstrate
    improvements over existing data-driven and classical fingerprinting
    benchmarks, robustness to degraded inference-time DT information
    without retraining, and improved empirical proposal uncertainty calibration
    as the number of APs increases.
\end{itemize}

\subsubsection*{Organization} The remainder of this paper is organized as follows. 
Section~\ref{sec:system_model} presents the system model and formulates wireless localization as a DT-guided probabilistic inverse problem. 
Section~\ref{sec:proposed_framework} develops the proposed probabilistic inverse learning framework. Section~\ref{sec:implementation} summarizes the practical implementation of the proposed scheme, and the benchmark schemes. 
Section~\ref{section:numerical_results} presents numerical experiments in site-specific DT environments and assesses the proposed scheme against classical fingerprinting and data-driven baselines. Section~\ref{section:conclusion} concludes the paper.

\subsubsection*{{Notation}}

Scalars, vectors, and matrices are denoted by lowercase letters, bold lowercase letters, and bold uppercase letters, respectively. 
The sets of real and complex numbers are denoted by $\mathbb R$ and $\mathbb C$. 
For a vector $\boldsymbol x$, $\|\boldsymbol x\|_2$ denotes the Euclidean norm, while $\Re\{\boldsymbol x\}$ and $\Im\{\boldsymbol x\}$ denote its real and imaginary components. The expectation is denoted by $\mathbb E[\cdot]$, and $\operatorname{diag}(\cdot)$ denotes a diagonal matrix.

\section{System Model and Problem Formulation}
\label{sec:system_model}

This section introduces the multi-AP channel observation model, the site-specific digital-twin (DT) forward model, and the resulting DT-informed probabilistic localization problem.

\subsection{Multi-AP Channel Observation Model}
\label{subsec:channel_observation_model}

We consider an uplink localization system with $N$ distributed single-antenna access points (APs) deployed at known positions in a site-specific propagation environment. A UE is located at an unknown position $\boldsymbol u\in\Omega\subset\mathbb R^{d_u}$, where $d_u$ denotes the localization dimension. The localization input is the observed multi-AP CIR feature associated with this UE position.

For AP $n\in \mathcal{N} = \{ 1,\ldots, N\}$, let
$\tilde{\boldsymbol h}_n(\boldsymbol u)\in\mathbb C^{K}$
denote the sampled delay-domain CIR associated with UE location
$\boldsymbol u$, where $K$ is the number of retained CIR taps.
The CIRs from all APs are stacked as $\tilde{\boldsymbol H}(\boldsymbol u)
= \big[\tilde{\boldsymbol h}_1^\top(\boldsymbol u), \ldots,
\tilde{\boldsymbol h}_N^\top(\boldsymbol u) \big]^\top \in\mathbb C^{NK}.$
The corresponding nominal real-valued CIR feature is obtained by
concatenating its real and imaginary components: $\boldsymbol y^{(0)}(\boldsymbol u)
=
\big[
\Re\{\tilde{\boldsymbol H}(\boldsymbol u)\}^{\top},
\Im\{\tilde{\boldsymbol H}(\boldsymbol u)\}^{\top}
\big]^{\top}
\in\mathbb R^{D_y}$, where $D_y=2NK$. For compactness, we denote by
\begin{align*}
\mathcal G:\Omega\rightarrow\mathbb R^{D_y},
\qquad
\boldsymbol y^{(0)}(\boldsymbol u)=\mathcal G(\boldsymbol u),
\end{align*}
the nominal site-specific mapping from UE position to the processed multi-AP CIR feature, including the adopted feature construction and normalization. The observed CIR feature is
\begin{align}
\boldsymbol y_{\mathrm{obs}}
=
\mathcal G(\boldsymbol u)
+
\boldsymbol e,
\label{eq:observation_model}
\end{align}
where $\boldsymbol e\in\mathbb R^{D_y}$ captures effective observation-domain errors, including channel-estimation noise, receiver noise, and preprocessing mismatch in the processed CIR feature.

\subsection{Site-Specific Digital Twin Forward Model}
\label{subsec:dt_forward_model}

The DT provides an approximate forward operator
\begin{align*}
\hat{\mathcal G}:\Omega\rightarrow\mathbb R^{D_y},
\end{align*}
which maps a hypothesized UE location to a predicted processed CIR feature in the same feature space as $\boldsymbol y_{\mathrm{obs}}$. For a candidate location $\boldsymbol v\in\Omega$, the DT ray tracer generates propagation paths using the site geometry, AP locations, and modeled electromagnetic interactions. The resulting channel response is processed using the same feature construction as the observed CIR feature, yielding $\hat{\mathcal G}(\boldsymbol v)$.

The DT is not assumed to be an exact replica of the physical environment. In practice, $\hat{\mathcal G}(\cdot)$ may differ from $\mathcal G(\cdot)$ due to geometric inaccuracies, material uncertainty, calibration errors, finite ray-tracing resolution, and unmodeled propagation effects. Moreover, the proposed formulation does not require $\hat{\mathcal G}(\cdot)$ to be differentiable with respect to UE location. We therefore treat $\hat{\mathcal G}(\cdot)$ as an approximate site-specific forward consistency model, rather than as an exact inverse solver or a differentiable trainable layer.

\subsection{DT-Informed Probabilistic Localization}
\label{subsec:localization_problem}

The localization task is to estimate the unknown UE position
$\boldsymbol u\in\Omega$ from the observed channel measurement
$\boldsymbol y_{\mathrm{obs}}$. From \eqref{eq:observation_model}, this is an
inverse problem associated with the forward mapping $\mathcal G(\cdot)$.
A conventional learning-based localization scheme uses a deterministic
estimator $\mathcal E:\mathbb R^{D_y}\rightarrow\Omega$, trained offline to
map each processed channel observation to a point estimate. The corresponding
minimum mean-squared error design criterion is
\begin{align}
\mathcal E^\star
=
\arg\min_{\mathcal E\in\mathcal X}
\mathbb E_{\boldsymbol u,\boldsymbol e}
\Big[
\Big\|
\boldsymbol u
- \mathcal E\!\big(
\underbrace{\mathcal G(\boldsymbol u)+\boldsymbol e}_{\boldsymbol y_{\mathrm{obs}}}
\big)
\Big\|_2^2
\Big],
\tag{P1}
\label{eq:inverse_localization_objective}
\end{align}
where $\mathcal X$ denotes the class of admissible deterministic estimators,
and the expectation is taken over the UE location distribution and the
observation residual in \eqref{eq:observation_model}.

However, the objective in
\eqref{eq:inverse_localization_objective} learns an observation-to-location
mapping without providing an explicit mechanism for evaluating whether a
hypothesized location is consistent with the site-specific propagation
response. Since the DT provides a forward mapping from a candidate location
to a predicted channel feature, it can provide such a consistency check. We
therefore define the DT consistency mismatch of a candidate location
$\boldsymbol v\in\Omega$ as
\begin{align}
\ell_{\mathrm{DT}}
\left(\boldsymbol v;\boldsymbol y_{\mathrm{obs}}\right)
=
d_{\mathrm{ch}}\left(
\boldsymbol y_{\mathrm{obs}},
\hat{\mathcal G}(\boldsymbol v)
\right),
\label{eq:dt_mismatch}
\end{align}
where $d_{\mathrm{ch}}(\cdot,\cdot)$ measures the discrepancy between
the observed and DT-predicted channel features, with a smaller
$\ell_{\mathrm{DT}}$ indicating stronger consistency. This mismatch provides
the site-specific forward-consistency information needed to compare
hypothesized UE locations. However, it does not by itself specify how the
inverse search over $\Omega$ should be performed. In particular, multipath
can make multiple locations plausible from the observation, while evaluating
the DT over the complete deployment domain would defeat the purpose of an
efficient learned inverse model. We therefore introduce a learned
conditional proposal distribution that restricts DT evaluation to a
data-informed set of plausible candidate locations.

Let
$\{\Pi_\theta(\cdot\mid\boldsymbol y_{\mathrm{obs}}):\theta\in\Theta\}$
denote a family of learned conditional proposal distributions for candidate
UE locations. Given the observed CIR feature,
$\Pi_\theta(\cdot\mid\boldsymbol y_{\mathrm{obs}})$ defines a sampling
distribution over candidate locations.

The proposal must satisfy two complementary requirements. First, it
should assign high likelihood to the ground-truth UE location so that the
candidate set remains anchored to the observation-driven localization
information. Second, it should retain a suitable spatial spread so that
finite candidate sampling provides useful alternatives for DT-guided
refinement. An overly concentrated proposal may fail to include corrective
candidates when its nominal estimate is inaccurate, whereas an excessively
diffuse proposal reduces the efficiency of finite candidate sampling. To
encourage candidate generation that remains useful for subsequent
DT-consistency scoring, we introduce the proposal regularization functional
\begin{align}
\mathcal R_{\mathrm{prop}}
\!\left(
\Pi_\theta(\cdot\mid\boldsymbol y_{\mathrm{obs}});
\boldsymbol y_{\mathrm{obs}}
\right),
\label{eq:proposal_regularizer}
\end{align}
whose concrete finite-candidate realization is developed in
Section~\ref{sec:proposed_framework}, where locations sampled from the
proposal are evaluated using the DT mismatch in
\eqref{eq:dt_mismatch}.

Combining these two complementary requirements, the probabilistic
localization objective is written as

\begin{align}
\begin{split}
\theta^\star
=
\arg\min_{\theta\in\Theta}
\;
&\mathbb E_{\boldsymbol u,\boldsymbol e}
\!\left[
-
\log
\Pi_\theta
(
\boldsymbol u
\mid
\boldsymbol y_{\mathrm{obs}}
)
\right]
\\ &+
\lambda
\,
\mathbb E_{\boldsymbol u,\boldsymbol e}
\!\left[
\mathcal R_{\mathrm{prop}}
\left(
\Pi_\theta(\cdot\mid\boldsymbol y_{\mathrm{obs}});
\boldsymbol y_{\mathrm{obs}}
\right)
\right].
\end{split}
\tag{P2}
\label{eq:prob_inverse_problem}
\end{align}

The first term is a supervised probabilistic localization objective
that encourages the proposal to assign high likelihood to the labeled UE
location. The second term shapes the proposal used for candidate generation
so that the resulting finite candidate set is suitable for DT-guided
refinement, while $\lambda\geq 0$ controls the relative contribution of the
two terms. During inference, the DT mismatch in
\eqref{eq:dt_mismatch} is likewise used to score and reweight candidates
sampled from the learned proposal.

\section{Proposed Probabilistic Inverse Learning Framework}
\label{sec:proposed_framework}

\begin{figure*}[t!]
\centering
\resizebox{!}{0.4\textheight}{
\begin{tikzpicture}[
    x=1cm,y=1cm,
    font=\sffamily\small,
    >=Stealth,
    panel/.style={draw=gray!55, rounded corners=7pt, thick, fill=gray!2},
    subpanel/.style={draw=gray!45, rounded corners=5pt, thick, fill=white},
    block/.style={draw=gray!65, rounded corners=5pt, thick, fill=white,
                  align=center, minimum height=1.15cm},
    learned/.style={draw=blue!65!black, rounded corners=5pt, thick, fill=blue!5,
                  align=center, minimum height=1.15cm},
    dtblock/.style={draw=teal!60!black, rounded corners=5pt, thick, fill=teal!6,
                  align=center, minimum height=1.15cm},
    outblock/.style={draw=orange!75!black, rounded corners=5pt, thick, fill=orange!8,
                  align=center, minimum height=1.15cm},
    title/.style={font=\bfseries\large, align=left},
    subtitle/.style={font=\bfseries\normalsize, align=center},
    note/.style={font=\scriptsize, align=center},
    smallnote/.style={font=\footnotesize, align=center},
    arrow/.style={->, very thick, draw=black!80},
    softarrow/.style={->, thick, draw=gray!70},
    feedback/.style={->, dashed, very thick, draw=blue!70!black},
    ap/.style={circle, draw=black!80, fill=white, thick, inner sep=1.2pt},
    ue/.style={circle, draw=black!80, fill=orange!75, thick, inner sep=2.0pt},
    ray/.style={dashed, thick, draw=blue!70!black},
    reflray/.style={dotted, thick, draw=teal!70!black}
]
\draw[panel] (0,5.95) rectangle (19.2,12.10);
\node[title, anchor=west] at (0.65,11.68)
{(a) Proposed probabilistic localization framework};

\node[subtitle, color=blue!70!black] at (7.9,10.95)
{I. Neural inference and candidate generation};

\node[block, text width=2.75cm] (obs) at (2.3,9.65)
{\textbf{Observed Feature}\\[1pt]
multi-AP CIR\\[1pt]
$\boldsymbol y_{\mathrm{obs}}$};

\node[learned, text width=3.45cm] (tf) at (6.8,9.65)
{\textbf{Transformer-based}\\
\textbf{Inverse Model}\\[1pt]
channel-to-location\\
representation};

\node[learned, text width=3.35cm] (prop) at (11.45,9.65)
{\textbf{Neural Proposal}\\[1pt]
$q(\boldsymbol v \mid \boldsymbol y_{\mathrm{obs}};\boldsymbol W)$\\[1pt]
outputs $\boldsymbol\mu,\boldsymbol\Sigma$};

\node[block, text width=3.10cm] (samp) at (16.05,9.65)
{\textbf{Candidate Sampling}\\[1pt]
sample plausible points\\[1pt]
$\{\boldsymbol v_i\}_{i=1}^{S}$};

\draw[arrow] (obs.east) -- (tf.west);
\draw[arrow] (tf.east) -- (prop.west);
\draw[arrow] (prop.east) -- (samp.west);

\node[subtitle, color=teal!70!black] at (7.9,7.95)
{II. DT-guided consistency scoring and refinement};

\node[block, text width=2.75cm, fill=orange!5] (uhat) at (2.3,6.75)
{\textbf{Final Estimate}\\[3pt]
$\hat{\boldsymbol u}$};

\node[outblock, text width=3.15cm] (comb) at (6.8,6.75)
{\textbf{Combine Candidates}\\[1pt]
weighted sum of\\
candidates};

\node[dtblock, text width=3.40cm] (score) at (11.45,6.75)
{\textbf{DT Consistency Scoring}\\[1pt]
compare $\boldsymbol y_{\mathrm{obs}}$ with\\
$\hat{\mathcal G}(\boldsymbol v_i)$, weights $w_i$};

\node[dtblock, text width=3.00cm] (dt) at (16.05,6.75)
{\textbf{Site-Specific DT}\\[1pt]
$\hat{\mathcal G}(\cdot)$\\
predicts features};

\draw[arrow] (samp.south) -- ++(0,-0.55) -- (dt.north);
\draw[arrow] (dt.west) -- (score.east);
\draw[arrow] (score.west) -- (comb.east);
\draw[arrow] (comb.west) -- (uhat.east);

\draw[feedback] (score.north) -- ++(0,0.55) -| (prop.south);
\node[note, text=blue!70!black, fill=gray!2, inner sep=1pt]
    at (12.10,8.55) {Proposal regularization during training};

\begin{scope}[yshift=15.80cm]
\draw[panel] (0,-18.85) rectangle (19.2,-10.15);
\node[title, anchor=west] at (0.70,-10.65) {(b) Algorithmic execution layout: Offline training vs. online inference};

\draw[gray!40, dashed, line width=1.5pt] (5.5, -11.2) -- (5.5, -18.6);
\node[subtitle, rotate=90, color=gray!50] at (5.1, -16.9) {Domain Boundary};
\node[subtitle, color=gray!50] at (2.7, -18.5) {Physical Domain};
\node[subtitle, color=gray!50] at (12.5, -18.5) {Digital / Neural Domain};

\node[subtitle, color=blue!75!black, anchor=west] at (0.7, -11.6) {OFFLINE TRAINING PHASE};
\node[subtitle, color=teal!75!black, anchor=west] at (0.7, -14.6) {ONLINE INFERENCE PHASE (DT used for candidate refinement)};

\node[block, text width=2.4cm] (train_obs) at (2.4, -13.1) {\textbf{Training Dataset}\\$\{ \boldsymbol u^{(m)},\boldsymbol y^{(m)}\}$};
\node[learned, text width=3.0cm] (train_tf) at (7.2, -13.1) {\textbf{Inverse Model}\\(Transformer)};
\node[learned, text width=3.0cm] (train_prop) at (12.2, -13.1) {\textbf{Probabilistic}\\ \textbf{Proposal } $q(\cdot)$};
\node[dtblock, text width=2.8cm] (train_cons) at (16.8, -13.1) {\textbf{Proposal Regularization}};

\draw[arrow] (train_obs.east) -- (train_tf.west);
\draw[arrow] (train_tf.east) -- (train_prop.west);
\draw[arrow] (train_prop.east) -- (train_cons.west);

\draw[feedback] (train_cons.north) -- ++(0,0.6) -| (train_tf.north);

\node[block, text width=2.4cm] (inf_obs) at (2.4, -17.4) {\textbf{Live Channel}\\ \textbf{Observation } $\boldsymbol y_{\mathrm{obs}}$};
\node[learned, text width=3.0cm, fill=blue!1] (inf_tf) at (7.2, -17.4) {\textbf{Deployed model weights}};
\node[learned, text width=3.0cm, fill=blue!1] (inf_prop) at (12.2, -17.4) {\textbf{Online Proposal}\\ $q(\boldsymbol v \mid \boldsymbol y_{\mathrm{obs}})$};
\node[outblock, text width=2.8cm] (inf_est) at (16.8, -17.4) {\textbf{DT-Refined}\\ \textbf{Estimate } $\hat{\boldsymbol u}$};

\draw[arrow] (inf_obs.east) -- (inf_tf.west);
\draw[arrow] (inf_tf.east) -- (inf_prop.west);
\draw[arrow] (inf_prop.east) -- (inf_est.west);

\draw[->, line width=1.2pt, dashed] (train_tf.south) -- (inf_tf.north);
\node[right, font=\scriptsize\bfseries] at (7.3, -15.25) {Deploy Parametric Weights};

\node[dtblock, fill=teal!15, text width=2.8cm, draw=teal, line width=1.2pt] (shared_dt) at (16.8, -15.25) {\textbf{Site-Specific DT}\\ Forward Model $\hat{\mathcal{G}}(\cdot)$};
\draw[softarrow, teal] (shared_dt.north) -- (train_cons.south);
\draw[softarrow, teal] (shared_dt.south) -- (inf_est.north);
\end{scope}
\end{tikzpicture}%
}
\caption{Overview of \emph{PRISM}. (a) The Transformer-based inverse model maps the observed channel
feature to a neural proposal distribution
$q(\boldsymbol v\mid\boldsymbol y_{\mathrm{obs}};\boldsymbol W)$, from which
candidate UE locations are sampled. The DT then
predicts the corresponding channel features and assigns consistency weights $w_i$, which are used to reweight sampled candidates and refine the 
localization estimate $\hat{\boldsymbol u}$. (b) During offline training, the proposal is learned from labeled data and regularized for finite-candidate generation. During online inference, DT consistency explicitly reweights sampled candidates to refine the location estimate.}
\label{fig:proposed_architecture}
\end{figure*}
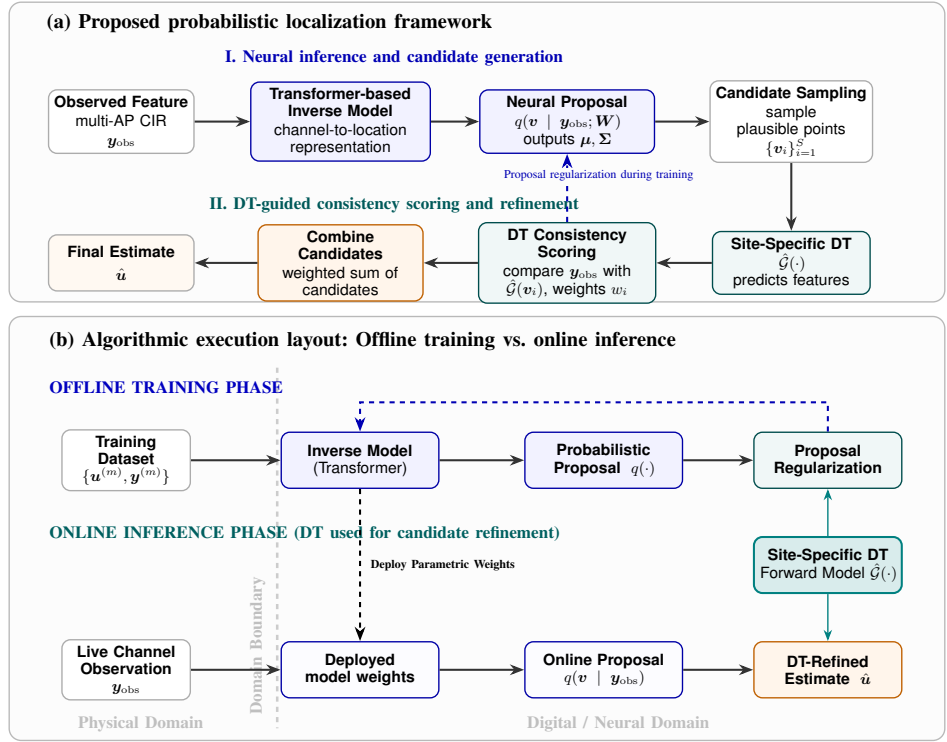

This section develops the finite-candidate learning and inference framework
associated with Problem~\eqref{eq:prob_inverse_problem}. A neural inverse model
learns a conditional proposal distribution over UE locations, with training
regularization used to maintain a candidate region suitable for subsequent
refinement. At inference, candidates sampled from this proposal are evaluated
using the DT mismatch in \eqref{eq:dt_mismatch} and reweighted to obtain the
refined location estimate. An overview is shown in
Fig.~\ref{fig:proposed_architecture}.

\subsection{Neural Inverse Proposal}
\label{subsec:neural_inverse_proposal}

The proposal distribution in \eqref{eq:prob_inverse_problem} is represented
in the proposed framework by a conditional distribution
$q(\cdot\mid\boldsymbol y_{\mathrm{obs}};\boldsymbol W)$. The weights
$\boldsymbol W$ parameterize the neural mapping from the observed CIR
feature to the parameters of $q$. In this work, we use a Gaussian proposal
distribution to obtain a tractable parameterization of the probabilistic
localization output,
\begin{align}
q(\boldsymbol u\mid\boldsymbol y_{\mathrm{obs}};\boldsymbol W)
=
\mathcal N
\left(
\boldsymbol u;
\boldsymbol\mu(\boldsymbol y_{\mathrm{obs}};\boldsymbol W),
\boldsymbol\Sigma(\boldsymbol y_{\mathrm{obs}};\boldsymbol W)
\right),
\label{eq:proposal_distribution}
\end{align}
where $\boldsymbol\mu(\boldsymbol y_{\mathrm{obs}};\boldsymbol W)$ is the
predicted location mean and
$\boldsymbol\Sigma(\boldsymbol y_{\mathrm{obs}};\boldsymbol W)$ is the
predicted covariance.

The mean provides the nominal observation-driven localization estimate,
while the covariance determines the spatial region covered by candidate
sampling. A concentrated proposal focuses candidate generation near the
nominal estimate, whereas a broader proposal provides a larger region in
which DT-guided refinement can identify alternative locations. The proposal
parameters therefore determine both the nominal localization output and the
candidate region available for subsequent DT consistency evaluation.

Let
$\mathcal D=\{(\boldsymbol u^{(m)},\boldsymbol y^{(m)})\}_{m=1}^{M}$
be a given training set. The supervised term is the empirical negative
log-likelihood of the ground-truth UE location under the proposal,
\begin{align}
L_{\mathrm{sup}}^{(m)}
=
-
\log q
\left(
\boldsymbol u^{(m)}
\mid
\boldsymbol y^{(m)};
\boldsymbol W
\right).
\label{eq:supervised_loss_compact}
\end{align}
For the Gaussian proposal in \eqref{eq:proposal_distribution}, this becomes
\begin{align}
L_{\mathrm{sup}}^{(m)}
&=
\frac{d_u}{2}\log(2\pi)
+
\frac12
\log\det
\left(
\boldsymbol\Sigma^{(m)}
\right)
\label{eq:supervised_loss}
\\
&\quad+
\frac12
\left(
\boldsymbol u^{(m)}-\boldsymbol\mu^{(m)}
\right)^\top
\left(
\boldsymbol\Sigma^{(m)}
\right)^{-1}
\left(
\boldsymbol u^{(m)}-\boldsymbol\mu^{(m)}
\right).
\nonumber
\end{align}
Here,
$\boldsymbol\mu^{(m)}
=
\boldsymbol\mu(\boldsymbol y^{(m)};\boldsymbol W)$,
$\boldsymbol\Sigma^{(m)}
=
\boldsymbol\Sigma(\boldsymbol y^{(m)};\boldsymbol W)$,
and $d_u$ is the localization dimension.

\subsection{DT-Based Consistency Weighting}
\label{subsec:dt_consistency_weighting}

The DT mismatch in \eqref{eq:dt_mismatch} provides the
forward-consistency criterion used to compare candidate locations. For
notational convenience, we denote the corresponding DT consistency energy
by
\begin{align}
E(\boldsymbol v;\boldsymbol y_{\mathrm{obs}})
\triangleq
\ell_{\mathrm{DT}}
\left(
\boldsymbol v;
\boldsymbol y_{\mathrm{obs}}
\right)
=
d_{\mathrm{ch}}
\left(
\boldsymbol y_{\mathrm{obs}},
\hat{\mathcal G}(\boldsymbol v)
\right),
\label{eq:reconstruction_energy}
\end{align}
where a lower value of
$E(\boldsymbol v;\boldsymbol y_{\mathrm{obs}})$ indicates stronger
agreement between the observed CIR feature and the channel feature predicted
by the DT at candidate location $\boldsymbol v$.

Rather than evaluating this consistency criterion over the complete
deployment domain, we apply it to a finite set of locations generated by the
learned proposal. For a given observation
$\boldsymbol y_{\mathrm{obs}}$, $S$ candidate locations are sampled
independently according to
\begin{align}
\boldsymbol v_i
\sim
q(\boldsymbol v\mid\boldsymbol y_{\mathrm{obs}};\boldsymbol W),
\qquad
i=1,\ldots,S.
\label{eq:proposal_samples}
\end{align}
Each candidate is evaluated using
\begin{align}
E_i
=
E(\boldsymbol v_i;\boldsymbol y_{\mathrm{obs}}),
\label{eq:sample_energy}
\end{align}
and assigned the normalized DT-consistency weight
\begin{align}
w_i
=
\frac{
\exp\left(-E_i/\tau\right)
}{
\sum_{j=1}^{S}
\exp\left(-E_j/\tau\right)
},
\label{eq:normalized_weights}
\end{align}
where $\tau>0$ controls the sensitivity of the weights to the DT mismatch.

The normalization in \eqref{eq:normalized_weights} converts the
candidate-wise DT mismatches into relative consistency weights over the
finite sampled set. Candidates with smaller mismatch receive larger weight,
while decreasing $\tau$ concentrates the weighting more strongly on the
most DT-consistent candidates. Thus,
$\{(\boldsymbol v_i,w_i)\}_{i=1}^{S}$ represents the relative preference
supplied by the DT among the locations proposed for a given observation.

\subsection{Proposal Regularization}
\label{subsec:proposal_regularization}

We now instantiate the proposal regularization functional in
Problem~\eqref{eq:prob_inverse_problem} using the DT-weighted finite
candidate set. For each training observation, candidate locations are
sampled from the current proposal and assigned relative weights according
to their DT consistency. These weighted samples provide a finite-candidate
mechanism for incorporating site-specific forward-consistency information
into the proposal-learning objective.

For the $m$th training pair
$(\boldsymbol u^{(m)},\boldsymbol y^{(m)})$, candidate locations
$\{\boldsymbol v_i^{(m)}\}_{i=1}^{S}$ are sampled from
$q(\cdot\mid\boldsymbol y^{(m)};\boldsymbol W)$ and evaluated using
\begin{align}
E_i^{(m)}
=
E
\left(
\boldsymbol v_i^{(m)};
\boldsymbol y^{(m)}
\right).
\label{eq:training_sample_energy}
\end{align}
The normalized DT-consistency weights can then be computed using \eqref{eq:normalized_weights}. Subsequently, using these relative DT-consistency weights, we define the
finite-candidate proposal regularization term as
\begin{align}
L_{\mathrm{prop}}^{(m)}
=
-
\sum_{i=1}^{S}
w_i^{(m)}
\log q
\left(
\boldsymbol v_i^{(m)}
\mid
\boldsymbol y^{(m)};
\boldsymbol W
\right).
\label{eq:proposal_loss}
\end{align}

Equation~\eqref{eq:proposal_loss} gives greater contribution to
candidate locations that exhibit stronger agreement with the DT-predicted
channel response. The regularizer therefore incorporates the relative
preference provided by the DT over samples generated by the current
proposal. It is a finite-candidate proposal regularizer rather than an exact
likelihood, posterior, or distribution-matching objective over the complete
localization domain. Importantly, evaluating
\eqref{eq:proposal_loss} requires only the numerical DT consistency scores
associated with the sampled candidates and does not require derivatives of
the DT forward model $\hat{\mathcal G}(\cdot)$.

With this, the overall training objective is
\begin{align}
\min_{\boldsymbol W}
\quad
\frac{1}{M}
\sum_{m=1}^{M}
\left(
L_{\mathrm{sup}}^{(m)}
+
\lambda
L_{\mathrm{prop}}^{(m)}
\right),
\label{eq:training_objective}
\end{align}
where $\lambda\geq0$ controls the strength of the finite-candidate
proposal regularization. The supervised term anchors the learned proposal
to the labeled UE location, while the proposal-regularization term
incorporates relative DT consistency among sampled candidates. Together,
the two terms realize the complementary roles introduced in
Problem~\eqref{eq:prob_inverse_problem}: learning an observation-informed
location proposal while maintaining candidate generation that can benefit
from site-specific DT guidance.

\subsection{Closed-Loop Inference}
\label{subsec:closed_loop_inference}

At inference time, the trained inverse model first produces the
conditional proposal
$q(\boldsymbol u\mid\boldsymbol y_{\mathrm{obs}};\boldsymbol W)$.
Candidate locations are sampled according to
\eqref{eq:proposal_samples}, evaluated using the DT consistency energy in
\eqref{eq:sample_energy}, and assigned the relative consistency weights in
\eqref{eq:normalized_weights}. The DT therefore operates only on the finite
candidate region supplied by the learned inverse model rather than searching
over the complete deployment domain.

The final DT-refined location estimate is computed as the weighted candidate
mean
\begin{align}
\hat{\boldsymbol u}
=
\sum_{i=1}^{S}
w_i
\boldsymbol v_i,
\label{eq:weighted_mean_estimate}
\end{align}
where the weights $w_i$ are computed using
\eqref{eq:normalized_weights}.

The estimate in \eqref{eq:weighted_mean_estimate} is therefore
interpreted as a DT-consistency-weighted refinement over the sampled
candidate set, rather than as an exact posterior mean over $\Omega$. For a
small inference temperature $\tau$, the weights become concentrated on the
candidates with the smallest DT mismatch, and the estimator approaches a
soft best-candidate selection among the proposed locations.

The resulting framework closes the loop between inverse learning and
site-specific forward consistency. The learned inverse model restricts the
search to a data-informed candidate region, while the DT provides relative
consistency information that enters proposal regularization during training
and explicitly refines the final location estimate during inference.

\section{Localization architecture and Benchmark Schemes}
\label{sec:implementation}

This section summarizes the implementation details of the proposed PRISM localization scheme\footnote{\label{ft:code}Code can be found at https://github.com/AAU-CNT/cnt-isac-sionna/tree/twin-localization}. Moreover, we specify the learning-based and fingerprinting benchmark schemes used to evaluate its performance.

\subsection{Neural Proposal Architecture}
\label{subsec:neural_proposal_architecture}

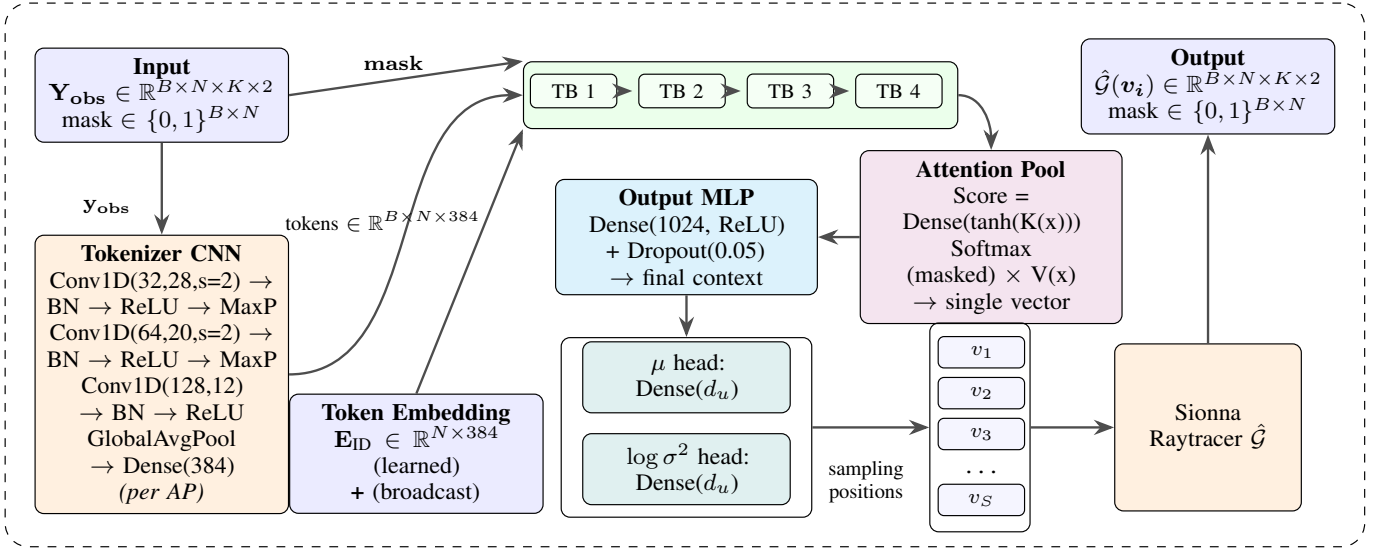
\begin{figure*}[t!]
\centering
\resizebox{\textwidth}{!}{

\begin{tikzpicture}[
    node distance=1.1cm and 0.9cm,
    every node/.style={font=\footnotesize},
    block/.style={rectangle, draw, rounded corners=3pt, minimum height=1.05cm,
                  text width=2.8cm, align=center, fill=blue!8},
    cnnblock/.style={rectangle, draw, rounded corners=3pt, minimum height=2.45cm,
                     text width=2.8cm, align=center, fill=orange!12},
    transcontainer/.style={rectangle, draw, rounded corners=3pt, 
                           minimum height=0.8cm, minimum width=5.2cm,
                           align=center, fill=green!8},
    miniblock/.style={rectangle, draw, rounded corners=2pt, minimum height=0.3cm,
                      text width=0.8cm, align=center, fill=green!4, font=\scriptsize},
  posblock/.style={rectangle, draw, rounded corners=2pt, minimum height=0.3cm,
                      text width=0.8cm, align=center, fill=blue!4, font=\scriptsize},
    attnblock/.style={rectangle, draw, rounded corners=3pt, minimum height=1.15cm,
                      text width=2.9cm, align=center, fill=purple!12},
    mlpblock/.style={rectangle, draw, rounded corners=3pt, minimum height=1.05cm,
                     text width=2.9cm, align=center, fill=cyan!12},
    headblock/.style={rectangle, draw, rounded corners=3pt, minimum height=0.85cm,
                      text width=2.25cm, align=center, fill=teal!12},
    distributioncontainer/.style={rectangle, draw, rounded corners=3pt, 
                           minimum height=2.15cm, minimum width=3cm,
                           align=center},
   poscontainer/.style={rectangle, draw, rounded corners=3pt, 
                           minimum height=2.5cm, minimum width=1.2cm,
                           align=center},
    sionna/.style={rectangle, draw, rounded corners=3pt, minimum height=2cm,
                     text width=2cm, align=center, fill=orange!12},
    arrow/.style={-{Stealth[scale=1.05]}, thick, color=black!70},
    dblarrow/.style={latex-latex, thick, color=purple!70},
    label/.style={font=\scriptsize, align=center},
    tensor/.style={font=\scriptsize\ttfamily, inner sep=1pt}
]


\node[block] (input) at (0,4.2) {
    \textbf{Input}\\
    $\mathbf{Y_{obs}} \in \mathbb{R}^{B \times N \times K \times 2}$\\
    mask $\in \{0,1\}^{B \times N}$
};

\node[cnnblock, below=of input] (cnn) {
    \textbf{Tokenizer CNN}\\
    Conv1D(32,28,s=2) $\to$ BN $\to$ ReLU $\to$ MaxP\\
    Conv1D(64,20,s=2) $\to$ BN $\to$ ReLU $\to$ MaxP\\
    Conv1D(128,12) $\to$ BN $\to$ ReLU\\
    GlobalAvgPool $\to$ Dense(384)\\
    \emph{(per AP)}
};

\node[block, right=1.0cm of cnn, yshift=-0.95cm, xshift=-1cm] (idemb) {
    \textbf{Token Embedding}\\
    $\mathbf{E}_{\text{ID}} \in \mathbb{R}^{N \times 384}$ (learned)\\
    \textbf{+} (broadcast)
};


\node[transcontainer, right=2.8cm of input] (tx) {
};

\begin{scope}[shift={(tx.center)}, xshift=-2cm, yshift=0.05cm]
    \node[miniblock] (tb1) at (0,0) {TB 1};
    \node[miniblock, right=0.25cm of tb1] (tb2) {TB 2};
    \node[miniblock, right=0.25cm of tb2] (tb3) {TB 3};
    \node[miniblock, right=0.25cm of tb3] (tb4) {TB 4};
\end{scope}

\draw[arrow, shorten >=2pt, shorten <=2pt] (tb1.east) -- (tb2.west);
\draw[arrow, shorten >=2pt, shorten <=2pt] (tb2.east) -- (tb3.west);
\draw[arrow, shorten >=2pt, shorten <=2pt] (tb3.east) -- (tb4.west);


\node[attnblock, below=0.25cm of tx, xshift=+3cm] (attnpool) {
    \textbf{Attention Pool}\\
    Score = Dense(tanh(K(x)))\\
    Softmax (masked) $\times$ V(x)\\
    $\to$ single vector
};

\node[mlpblock, left=0.5cm of attnpool] (mlp) {
    \textbf{Output MLP}\\
    Dense(1024, ReLU) + Dropout(0.05)\\
    $\to$ final context
};

\node[headblock, below=of mlp, yshift=+0.55cm] (mu) {
    $\mu$ head:\\ Dense($d_u$)
};
\node[headblock, below=of mlp, yshift=-0.55cm] (logvar) {
    $\log\sigma^2$ head:\\ Dense($d_u$)
};
\node[distributioncontainer, below=+0.5cm of mlp] (gauss) {
};
\node[poscontainer, right=+1.4cm of gauss] (posc) {
};
\node[posblock, right=+1.5cm of gauss, yshift=+0.9cm] (pos1) {
    $v_1$
};
\node[posblock, below=+0.1cm of pos1] (pos2) {
    $v_2$
};
\node[posblock, below=+0.1cm of pos2] (pos3) {
    $v_3$
};
\node[tensor, below=0.2cm of pos3] {...};
\node[posblock, below=+0.4cm of pos3] (posn) {
    $v_S$
};
\node[sionna, right=+1cm of posc] (RT) {
    Sionna Raytracer $\hat{\mathcal{G}}$
};

\node[block, above=2.5cm of RT] (Out)
{
    \textbf{Output}\\
    $\hat{\mathcal{G}}(\boldsymbol{v_i}) \in \mathbb{R}^{B \times N \times K \times 2}$\\
    mask $\in \{0,1\}^{B \times N}$
};

\draw[arrow] (input) -- (cnn);
\node[left=7pt, font=\scriptsize] 
     at ($(input)!0.4!(cnn)$) 
     {$\mathbf{y_{obs}}$};

\draw[arrow] (cnn.east) to[out=0,in=180] (tx.west);

\node[above=7pt, font=\scriptsize] 
     at ($(cnn.east)!0.4!(tx.west)$) 
     {tokens $\in \mathbb{R}^{B\times N\times384}$};

\draw[arrow] (tx.east) to[out=0,in=90] (attnpool.north);

\draw[arrow] (attnpool) -- (mlp);

\draw[arrow] (mlp.south) -- (gauss.north);

\draw[arrow] (idemb.north) -- (tx.south west);

\draw[arrow] (input.east) -- (tx.north west);
\node[above=5pt, font=\scriptsize] 
     at ($(input)!0.4!(tx)$) 
     {$\mathbf{mask}$};
\draw[arrow] (gauss.east) -- (posc.west);
\node[below=7pt, font=\scriptsize] 
     at ($(gauss.west)!0.65!(posc.east)$) 
     {$\mathbf{\begin{array}{c} \textrm{sampling} \\ \textrm{positions} \end{array}}$};
\draw[arrow] (posc) -- (RT);
\draw[arrow] (RT) -- (Out);

\begin{scope}[on background layer]
    \node[fit=(input)(attnpool)(logvar)(RT) (Out), draw, dashed, rounded corners=8pt,
          inner xsep=10pt, inner ysep=12pt, label=above:\textbf{APTransformerEncoder}] {};
\end{scope}

\end{tikzpicture}}
\caption{Architecture of the proposed probabilistic Transformer encoder. The multi-AP CIR input
$\mathbf Y_{\mathrm{obs}}\in\mathbb R^{B\times N\times K\times 2}$, together with an AP validity mask, is encoded by an AP-wise CNN tokenizer, augmented with learned AP-ID embeddings, processed by a masked Transformer encoder, and aggregated using attention pooling. The output context vector feeds two distributional heads that predict the mean $\boldsymbol\mu$ and log-variance $\log\boldsymbol\sigma^2$ of the neural proposal. Candidate locations sampled from this proposal are evaluated by the DT forward model $\hat{\mathcal G}(\cdot)$.}
\label{fig:proposed_transformer_architecture}
\end{figure*}

The neural proposal in \eqref{eq:proposal_distribution} is parameterized using the probabilistic Transformer encoder shown in Fig.~\ref{fig:proposed_transformer_architecture}. The input is the tensor form of the observed multi-AP CIR feature, $\mathbf Y_{\mathrm{obs}}\in\mathbb R^{B\times N\times K\times 2}$, where $B$ is the batch size, $N$ is the number of APs, $K$ is the CIR length, and the last dimension contains the real and imaginary components. A binary AP mask indicates which APs provide valid observations.

Each AP observation is encoded by a shared AP-wise tokenizer, producing one token per AP. A learned AP-ID embedding is added to preserve the identity of the AP associated with each token, since different APs observe different site-specific multipath responses. The resulting AP-token sequence is processed by a masked Transformer encoder, where the AP mask prevents unavailable APs from contributing to the attention operation. The encoded AP tokens are then aggregated by attention pooling into a single context vector, which is passed to distributional output heads that parameterize the neural proposal.

The output heads predict the proposal mean $\boldsymbol\mu(\boldsymbol y;\boldsymbol W)\in\mathbb R^{d_u}$ and the log-variance vector $\log\boldsymbol\sigma^2(\boldsymbol y;\boldsymbol W)\in\mathbb R^{d_u}$. Although \eqref{eq:proposal_distribution} allows a general covariance matrix, we use a diagonal covariance in the implementation,
\begin{align}
\boldsymbol\Sigma(\boldsymbol y;\boldsymbol W)
=
\operatorname{diag}
\left(
\sigma_1^2(\boldsymbol y;\boldsymbol W),
\ldots,
\sigma_{d_u}^2(\boldsymbol y;\boldsymbol W)
\right).
\label{eq:diagonal_covariance}
\end{align}
This diagonal Gaussian parameterization reduces the number of uncertainty parameters, avoids covariance-matrix inversion, and simplifies sampling. In preliminary trials for the scenarios considered in the numerical results, more expressive alternatives, including full-covariance and mixture-based proposals, did not provide any substantial gains. We therefore use the diagonal Gaussian proposal in all reported experiments.

Candidate locations are sampled from the neural proposal. For a given $\boldsymbol y$, samples are generated using the reparameterization
\begin{align}
\boldsymbol v_i
=
\boldsymbol\mu(\boldsymbol y;\boldsymbol W)
+
\boldsymbol\sigma(\boldsymbol y;\boldsymbol W)
\odot
\boldsymbol\xi_i,
\qquad
\boldsymbol\xi_i\sim\mathcal N(\boldsymbol 0,\mathbf I),
\label{eq:reparameterized_sampling}
\end{align}
where $\boldsymbol\sigma$ contains the predicted standard deviations and $\odot$ denotes element-wise multiplication.

\subsection{Training and DT-Guided Inference}
\label{subsec:training_inference}

For the diagonal covariance in \eqref{eq:diagonal_covariance}, omitting additive constants, we use the following supervised loss:
\begin{align}
L_{\mathrm{sup}}^{(m)}
=
\frac{\alpha}{2}
\sum_{d=1}^{d_u}
\log
\left(
(\sigma_d^{(m)})^2
\right)
+
\frac12
\sum_{d=1}^{d_u}
\frac{
\left(
u_d^{(m)}-\mu_d^{(m)}
\right)^2
}{
(\sigma_d^{(m)})^2
},
\label{eq:weighted_gaussian_nll}
\end{align}
where $0<\alpha\leq 1$ is the scaling constant of the log-variance term. Setting $\alpha=1$ recovers the standard diagonal Gaussian negative log-likelihood, while $\alpha<1$ relaxes the penalty on larger proposal variances. This provides an effective way to avoid overly concentrated proposals when DT-guided refinement benefits from a broader candidate set.

The DT consistency energy in \eqref{eq:sample_energy} is computed from a channel-feature similarity between the observed CIR feature $\boldsymbol y_{\mathrm{obs}}$ and the DT-predicted feature $\hat{\boldsymbol y}=\hat{\mathcal G}(\boldsymbol v)$. Specifically, for a normalized similarity score $S(\boldsymbol y_{\mathrm{obs}},\hat{\boldsymbol y})\in[0,1]$, we use
\begin{align}
E(\boldsymbol v;\boldsymbol y_{\mathrm{obs}})
=
1-S\left(
\boldsymbol y_{\mathrm{obs}},
\hat{\mathcal G}(\boldsymbol v)
\right).
\label{eq:energy_from_similarity}
\end{align}
We evaluate different choices of $S(\cdot,\cdot)$, namely the cosine similarity $S_{\mathrm{cos}}(\cdot,\cdot)$ and the structural similarity index $S_{\mathrm{SSIM}}(\cdot,\cdot)$. The cosine similarity between the element-wise magnitudes of the CIR features is defined as
\begin{align}
S_{\mathrm{cos}}
\left(
\boldsymbol y_{\mathrm{obs}},
\hat{\boldsymbol y}
\right)
=
\frac{
\sum_{\ell=1}^{D_y}
\left|
y_{\mathrm{obs},\ell}
\right|
\left|
\hat y_{\ell}
\right|
}{
\left\|
\boldsymbol y_{\mathrm{obs}}
\right\|_2
\left\|
\hat{\boldsymbol y}
\right\|_2
}.
\label{eq:cosine_sim}
\end{align}

The structural similarity index measure (SSIM) is computed by arranging the multi-AP CIR feature as a structured tensor over AP index, delay, and real/imaginary components. For local windows of the observed and DT-predicted CIR representations, SSIM is given by~\cite{ssim}
\begin{align}
S_{\mathrm{SSIM}}
=
\frac{
(2\mu_y\mu_{\hat y}+C_1)(2\sigma_{y\hat y}+C_2)
}{
(\mu_y^2+\mu_{\hat y}^2+C_1)
(\sigma_y^2+\sigma_{\hat y}^2+C_2)
},
\label{eq:ssim_similarity}
\end{align}
where $\mu_y$ and $\mu_{\hat y}$ are local means, $\sigma_y^2$ and $\sigma_{\hat y}^2$ are local variances, $\sigma_{y\hat y}$ is the local covariance, and $C_1,C_2$ are stabilizing constants. The resulting SSIM values are averaged over local windows to obtain the similarity score used in \eqref{eq:energy_from_similarity}. 

For PRISM, the similarity measure used for DT consistency scoring is
selected using validation performance. Among the considered measures,
$S_{\mathrm{SSIM}}$ provides the best validation performance and is
therefore used in the reported PRISM experiments.

At inference time, the trained Transformer predicts
$\boldsymbol\mu(\boldsymbol y_{\mathrm{obs}};\boldsymbol W)$ and
$\boldsymbol\Sigma(\boldsymbol y_{\mathrm{obs}};\boldsymbol W)$.
Candidate locations are sampled from the resulting proposal, evaluated using
the DT consistency energy, and assigned the normalized weights in
\eqref{eq:normalized_weights}. The final estimate is then obtained from the
weighted candidate mean in \eqref{eq:weighted_mean_estimate}.

\subsection{Benchmark Schemes}
\label{subsec:benchmarks}

We compare \emph{PRISM} with five benchmark localization schemes chosen to isolate the roles of neural architecture, probabilistic DT-guided refinement, and fingerprint database resolution. 

The first two benchmarks are direct neural predictors that map the observed CIR feature to a single UE location estimate: \emph{CNN-Direct}, a CNN-based model adapted from an existing neural fingerprinting architecture~\cite{CNN_fingerprinting_OG}, and \emph{Transformer-Direct}, a deterministic Transformer using the same encoder structure as the proposed proposal network but with a point-estimate output. 

Moreover, we include three fingerprinting benchmarks: \emph{C-FP}, a classical fingerprinting baseline, and \emph{Dense-FP}, a higher-resolution variant used to test whether \emph{PRISM} maintains its gains against a more computationally demanding database search. Finally, \emph{RFMap-FP}, based on~\cite{DT_localization_AK}, represents an RF-map fingerprinting variant within the broader family of classical database-matching localization schemes. For the classical fingerprinting schemes, we evaluated the considered channel-feature similarity measures on the validation set and found cosine similarity to provide the best performance; it is therefore used for \emph{C-FP} and \emph{Dense-FP}. We detail each benchmark scheme below.

\noindent \emph{1- CNN-Direct} is a direct neural localization benchmark based on a convolutional architecture adapted from~\cite{CNN_fingerprinting_OG}. It maps the observed CIR directly to a point estimate of the UE location,
\begin{align}
\hat{\boldsymbol u}_{\mathrm{CNN}}
= \mathcal E_{\mathrm{CNN}}
\left(
\boldsymbol y_{\mathrm{obs}}
\right).
\label{eq:cnn_baseline}
\end{align}
This benchmark is trained using labeled channel-location pairs and does not use DT-based consistency weighting or candidate refinement at inference.

\noindent \emph{2- Transformer-Direct} is a direct neural localization benchmark that uses the same encoder structure as the proposal network in \emph{PRISM}, but replaces the distributional output with a point-estimate head,
\begin{align}
\hat{\boldsymbol u}_{\mathrm{Tr}}
=
\mathcal E_{\mathrm{Tr}}
\left(
\boldsymbol y_{\mathrm{obs}}
\right).
\label{eq:transformer_baseline}
\end{align}
This benchmark isolates the effect of the Transformer-based CIR representation from the effect of probabilistic sampling and DT-guided refinement.

\noindent \emph{3- C-FP} is a classical fingerprinting benchmark using the validation-selected cosine similarity. Let
$\mathcal D_{\mathrm{FP}}
=\left\{
(\boldsymbol v_j,\boldsymbol y_j)
\right\}_{j=1}^{J}$
denote the fingerprint database, where $\boldsymbol v_j$ is a database location and $\boldsymbol y_j$ is the associated CIR fingerprint. Given an observation $\boldsymbol y_{\mathrm{obs}}$, fingerprinting selects the database location with maximum cosine similarity:
\begin{align}
j^\star
&=
\arg\max_{j\in\{1,\ldots,J\}}
S_{\mathrm{cos}}
\left(
\boldsymbol y_{\mathrm{obs}},
\boldsymbol y_j
\right), \, \hat{\boldsymbol u}_{\mathrm{FP}}
=
\boldsymbol v_{j^\star}.
\label{eq:fingerprinting_baseline}
\end{align}
This provides a classical database-matching benchmark.

\noindent \emph{4- Dense-FP} uses the same cosine-similarity rule as \emph{C-FP}, but with a denser fingerprint database. This benchmark provides a stronger fingerprinting comparison by increasing the spatial resolution of the database at the cost of a larger exhaustive search. It is used to test whether \emph{PRISM} maintains its gains against a more computationally demanding fingerprinting baseline with finer spatial resolution.

\noindent 5- \emph{RFMap-FP} is based on~\cite{DT_localization_AK} and represents an RF-map fingerprinting variant within the family of classical database-matching localization schemes. In the considered simulation setup, \emph{RFMap-FP} uses the same database as \emph{C-FP}, so the comparison primarily reflects the fingerprint matching rule. This benchmark provides a representative RF-map fingerprinting reference while remaining closely tied to \emph{C-FP}. 
\section{Numerical Experiments}
\label{section:numerical_results}

\subsection{Experimental Setup}
\label{subsec:experimental_setup}

\subsubsection{Scenario, Channel Generation, and AP Configurations}
\label{subsec:scenario_channel_ap}

The simulated channel data are generated using Sionna RT~\cite{Sionna_RT} in the Munich scene, which has been employed in several site-specific ray-tracing studies, e.g.,~\cite{munich_example,Munich_example_4,munich_trans,munich_OJ}. The scene represents an urban outdoor propagation environment. Additional experiments using the San Francisco scene provided with Sionna RT~\cite{Sionna_SF_scene} exhibited similar qualitative performance trends. The corresponding results are omitted here due to space limitations.

For channel generation, we use carrier frequency
$f_c=6\,\mathrm{GHz}$, bandwidth $B_{\mathrm{w}}=100\,\mathrm{MHz}$, and
frequency spacing $\Delta f=500\,\mathrm{kHz}$, yielding $200$
frequency samples. The corresponding CIR representation contains
$200$ delay taps over a $2\,\mu\mathrm{s}$ window with
$10\,\mathrm{ns}$ tap spacing.

All APs and UEs use single-element, vertically polarized dipole
antennas at a height of $1.15\,\mathrm{m}$. Unless otherwise stated, ray tracing uses $10^4$ rays per source, a maximum depth of $6$, and includes specular reflection, diffuse reflection, scattering, and refraction. The generated CIR features are kept in raw form followed by per-AP normalization.

\begin{figure}
\centering
\includegraphics[width=1\linewidth]{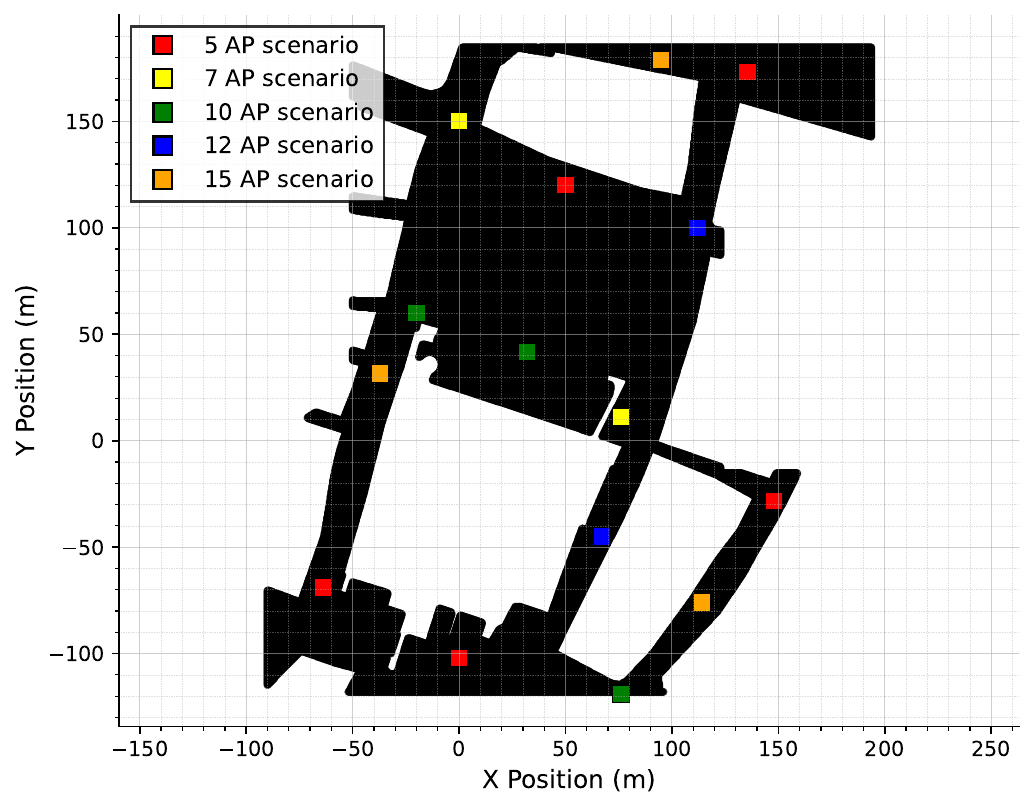}
\caption{2D layout of the Munich scene and AP positions. The shaded region indicates the candidate outdoor UE locations.}

\label{fig:position_mask}
\end{figure}

Five configurations with $N\in\{5,7,10,12,15\}$ distributed single-antenna APs are evaluated. Each configuration contains all AP positions used in the preceding smaller configuration, as shown in Fig.~\ref{fig:position_mask}, allowing the effect of increasing the AP count to be evaluated consistently. UE locations are placed at outdoor positions on a uniform grid with $1\,\mathrm{m}$ spacing along both spatial dimensions. This grid is used for training, validation, and testing of the learning-based schemes and as the default fingerprint database for \emph{C-FP}. For \emph{Dense-FP}, additional databases with $0.75\,\mathrm{m}$
and $0.5\,\mathrm{m}$ grid spacing are considered to control for the effect of fingerprint-database resolution when comparing against \emph{PRISM}.

\subsubsection{Training and Inference}
\label{subsec:training_inference_setup}

Each learning-based model is trained on the training set for $500$ epochs. After each epoch, the model is evaluated on the validation set, and the checkpoint with the lowest mean positional error is selected. Final performance is reported on a held-out test set.

All learning-based models use a batch size of $512$ and are trained with AdamW using a warm-up cosine learning-rate schedule. The warm-up fraction is $5\%$, the base learning rate is $5\times 10^{-4}$, and the minimum learning rate is $5\times 10^{-6}$.  For PRISM, we use the finite-candidate proposal regularization described in Section~\ref{subsec:proposal_regularization}. During inference, we use $S_{\mathrm{inf}}=200$ proposal samples and $\tau_{\mathrm{inf}}=10^{-4}$ for DT-guided candidate reweighting. The small inference temperature concentrates the estimate on the most DT-consistent candidates. The supervised loss uses the scaling factor $\alpha=0.5$, and the
proposal-regularization strength is set to $\lambda=1$. For the DT consistency energy, we use the SSIM-based similarity $S_{\mathrm{SSIM}}$ defined in \eqref{eq:ssim_similarity}.

\subsubsection{Observation Noise and DT-Feature Distortion}
\label{subsec:noise_dt_distortion}

We reserve the term SNR for additive noise applied to the observed CIR features. For a clean processed CIR feature $\boldsymbol y_i^{(0)}=\mathcal G(\boldsymbol u_i)$, the noisy observation is generated as $\boldsymbol y_i=\boldsymbol y_i^{(0)}+\boldsymbol e_i$, where $\boldsymbol e_i$ denotes additive observation noise. The observation SNR is
$$
\mathrm{SNR}_{\mathrm{obs}}
=
10\log_{10}
\left(
\frac{
\sum_{i=1}^{N_{\mathrm{obs}}}
\|\boldsymbol y_i^{(0)}\|_2^2
}{
\sum_{i=1}^{N_{\mathrm{obs}}}
\|\boldsymbol e_i\|_2^2
}
\right),
$$
where $N_{\mathrm{obs}}$ is the number of CIR observations used to set the noise level. Unless otherwise stated, experiments are conducted at $\mathrm{SNR}_{\mathrm{obs}}=25\,\mathrm{dB}$. This noise is applied to the observed CIR features used by the localization models and is distinct from imperfections in the DT forward model.

To evaluate robustness to imperfect inference-time DT information \emph{without retraining}, we distort the DT-predicted CIR features used for consistency scoring. For a candidate location $\boldsymbol v_i$, the distorted DT feature is modeled as $\hat{\mathcal G}_{\mathrm{dist}}(\boldsymbol v_i)=\hat{\mathcal G}(\boldsymbol v_i)+\boldsymbol\delta_{\mathrm{DT},i}$, where $\boldsymbol\delta_{\mathrm{DT},i}$ denotes the distortion applied to the DT-predicted feature. The distortion level is reported using the DT signal-to-distortion ratio
$$
\mathrm{SDR}_{\mathrm{DT}}
=
10\log_{10}
\left(
\frac{
\sum_{i=1}^{N_{\mathrm{DT}}}
\|\hat{\mathcal G}(\boldsymbol v_i)\|_2^2
}{
\sum_{i=1}^{N_{\mathrm{DT}}}
\|\boldsymbol\delta_{\mathrm{DT},i}\|_2^2
}
\right),
$$
where $N_{\mathrm{DT}}$ is the number of DT-predicted CIR features used to set the distortion level. Thus, $\mathrm{SDR}_{\mathrm{DT}}=25\,\mathrm{dB}$ has the same power-ratio interpretation as a $25\,\mathrm{dB}$ SNR, but refers to distortion of the DT-predicted features rather than noise in the observed CIRs.

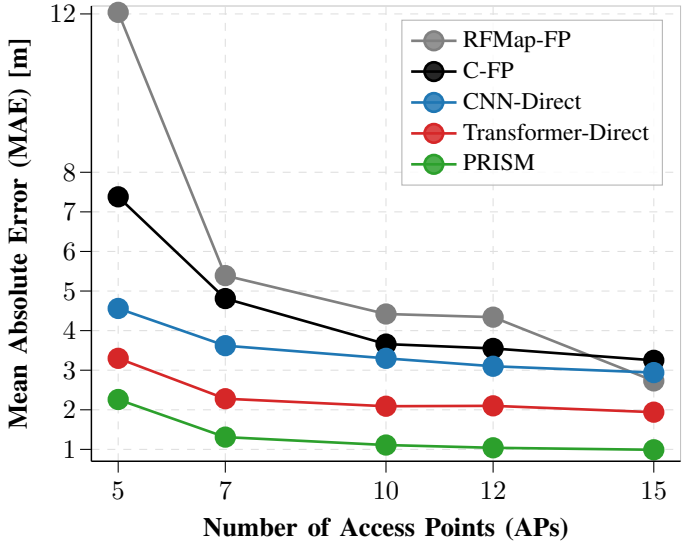
\begin{figure}[t]
\centering
\begin{tikzpicture}

\definecolor{crimson2143940}{RGB}{214,39,40}
\definecolor{darkgray176}{RGB}{176,176,176}
\definecolor{forestgreen4416044}{RGB}{44,160,44}
\definecolor{lightgray204}{RGB}{204,204,204}
\definecolor{steelblue31119180}{RGB}{31,119,180}

\begin{axis}[
grid style={dashed, gray!30},
width=0.88\columnwidth,
height=0.68\columnwidth,
scale only axis,
xlabel style={font=\bfseries, yshift=2pt},
ylabel style={
    font=\bfseries,
    at={(axis description cs:-0.075,0.5)},
    anchor=south
},
axis lines=box,
axis line style={line width=0.25pt, draw=black!25},
extra description/.code={
    \draw[black, line width=0.55pt]
        (rel axis cs:0,0) -- (rel axis cs:1,0);
    \draw[black, line width=0.55pt]
        (rel axis cs:0,0) -- (rel axis cs:0,1);
},
tick align=outside,
tick pos=left,
tick style={line width=0.4pt, draw=black},
xtick style={color=black},
ytick style={color=black},
yticklabel style={xshift=2pt},
xmajorgrids,
ymajorgrids,
x grid style={gray!25},
y grid style={gray!25},
legend cell align={left},
legend style={
    at={(0.97,0.97)},
    anchor=north east,
    font=\small,
    draw=black!30,
    fill=white,
    fill opacity=0.85,
    text opacity=1
},
xlabel={\textbf{Number of Access Points (APs)}},
xmin=4.5, xmax=15.5,
xtick={5,7,10,12,15},
ylabel={\textbf{Mean Absolute Error (MAE) [m]}},
ymin=0.7, ymax=12.2,
ytick={1,2,3,4,5,6,7,8,12}
]
\addplot [
    line width=1.0pt,
    gray,
    mark=*,
    mark size=3.5,
    mark options={solid}
]
coordinates {
    (5,12.04)
    (7,5.39)
    (10,4.42)
    (12,4.34)
    (15,2.73)
};
\addlegendentry{RFMap-FP}

\addplot [
    line width=1.0pt,
    black,
    mark=*,
    mark size=3.5,
    mark options={solid}
]
coordinates {
(5,7.38)
(7,4.81)
(10,3.66)
(12,3.55)
(15,3.25)
};
\addlegendentry{C-FP}

\addplot [
    line width=1.0pt,
    steelblue31119180,
    mark=*,
    mark size=3.5,
    mark options={solid}
]
coordinates {
(5,4.56)
(7,3.62)
(10,3.30)
(12,3.10)
(15,2.94)
};
\addlegendentry{CNN-Direct}

\addplot [
    line width=1.0pt,
    crimson2143940,
    mark=*,
    mark size=3.5,
    mark options={solid}
]
coordinates {
(5,3.30)
(7,2.28)
(10,2.09)
(12,2.10)
(15,1.94)
};
\addlegendentry{Transformer-Direct}

\addplot [
    line width=1.0pt,
    forestgreen4416044,
    mark=*,
    mark size=3.5,
    mark options={solid}
]
coordinates {
(5,2.26)
(7,1.31)
(10,1.11)
(12,1.04)
(15,0.99)
};
\addlegendentry{PRISM}

\end{axis}
\end{tikzpicture}
\caption{Mean absolute error versus number of APs for PRISM and benchmarks at an $\mathrm{SNR}_{\mathrm{obs}}$ of $25$ dB.}
\label{fig:mae_ap_PRISM_nn_fingerprinting}
\end{figure}

\subsection{Localization Accuracy and Distribution of Errors}
\label{subsec:ap_count_results}

\begin{figure}[t]
\centering
\makebox[\columnwidth][c]{%
\begin{tikzpicture}

\definecolor{darkgray176}{RGB}{176,176,176}
\definecolor{forestgreen4416044}{RGB}{44,160,44}
\definecolor{steelblue31119180}{RGB}{31,119,180}
\definecolor{orange25512714}{RGB}{255,127,14}

\begin{axis}[
grid style={dashed, gray!30},
width=0.88\columnwidth,
height=0.68\columnwidth,
scale only axis,
xlabel style={font=\bfseries, yshift=2pt},
ylabel style={
    font=\bfseries,
    at={(axis description cs:-0.075,0.5)},
    anchor=south
},
axis lines=box,
axis line style={line width=0.25pt, draw=black!25},
extra description/.code={
    \draw[black, line width=0.55pt]
        (rel axis cs:0,0) -- (rel axis cs:1,0);
    \draw[black, line width=0.55pt]
        (rel axis cs:0,0) -- (rel axis cs:0,1);
},
tick align=outside,
tick pos=left,
tick style={line width=0.4pt, draw=black},
xtick style={color=black},
ytick style={color=black},
yticklabel style={xshift=2pt},
xmajorgrids,
ymajorgrids,
x grid style={gray!25},
y grid style={gray!25},
legend cell align={left},
legend style={
    at={(0.97,0.97)},
    anchor=north east,
    font=\small,
    draw=black!30,
    fill=white,
    fill opacity=0.85,
    text opacity=1
},
xlabel={\textbf{Number of Access Points (APs)}},
xmajorgrids,
xmin=4.5, xmax=15.5,
xtick={5,7,10,12,15},
xtick style={color=black},
y grid style={darkgray176},
ylabel={\textbf{Mean Absolute Error [m]}},
ylabel style={
    at={(axis description cs:-0.075,0.5)},
    anchor=south
},
ymajorgrids,
ymin=0.7, ymax=8,
ytick={1,2,3,4,5,6,7,8},
yticklabels={1,2,3,4,5,6,7,8,9,10,11,12,13},
ytick style={color=black}
]

\addplot [
    line width=1.0pt,
    forestgreen4416044,
    mark=*,
    mark size=3.5,
    mark options={solid}
]
coordinates {
    (5,2.26)
    (7,1.31)
    (10,1.11)
    (12,1.04)
    (15,0.99)
};
\addlegendentry{PRISM (1 m)}

\addplot [
    line width=1.0pt,
    black,
    mark=*,
    mark size=3.5,
    mark options={solid}
]
coordinates {
    (5,7.38)
    (7,4.81)
    (10,3.66)
    (12,3.55)
    (15,3.25)
};
\addlegendentry{C-FP (1 m)}
\addplot [
    line width=1.0pt,
    cyan,
    mark=*,
    mark size=3.5,
    mark options={solid}
]
coordinates {
    (5,6.11)
    (7,3.81)
    (10,2.89)
    (12,2.64)
    (15,2.43)
};
\addlegendentry{Dense-FP (0.75 m)}
\addplot [
    line width=1.0pt,
    steelblue31119180,
    mark=*,
    mark size=3.5,
    mark options={solid}
]
coordinates {
    (5,4.88)
    (7,3.12)
    (10,2.17)
    (12,2.06)
    (15,1.94)
};
\addlegendentry{Dense-FP (0.5 m)}

\end{axis}
\end{tikzpicture}%
}
\caption{Mean absolute error versus number of APs for PRISM and classical fingerprinting methods for different resolutions.}
\label{fig:mae_ap_PRISM_fingerprinting_grids}
\end{figure}
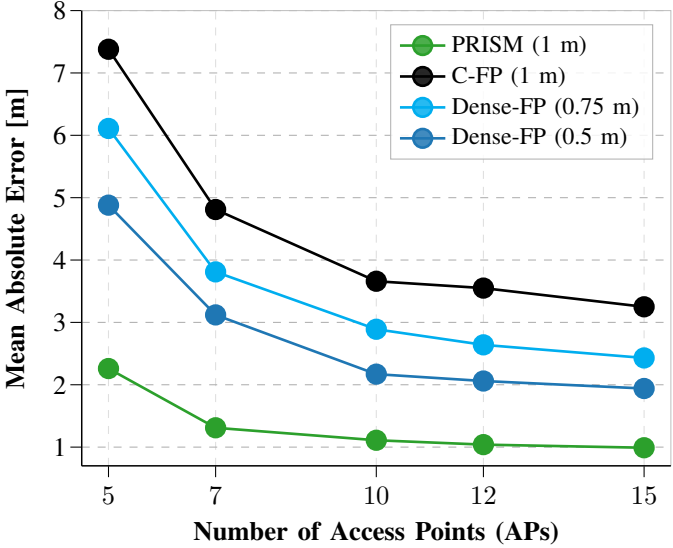

We first evaluate the impact of the number of APs on localization accuracy and quantify the gains provided by \emph{PRISM} over the learning-based and fingerprinting benchmarks. Fig.~\ref{fig:mae_ap_PRISM_nn_fingerprinting} shows that increasing the number of APs improves all schemes, with the largest reduction in MAE generally occurring from $5$ to $7$ APs. Beyond approximately $10$ APs, the improvements become more gradual, indicating diminishing returns from adding further APs. Among the direct learning-based benchmarks, \emph{Transformer-Direct} consistently outperforms \emph{CNN-Direct}, with MAEs of $2.09\,\mathrm{m}$ and $3.30\,\mathrm{m}$, respectively, at $10$ APs. This suggests that the Transformer architecture is better suited to aggregating the heterogeneous multi-AP CIR observations, as self-attention enables observation-dependent interactions across AP tokens. At the same $10$-AP configuration, \emph{PRISM} further reduces the MAE to $1.11\,\mathrm{m}$, corresponding to a $47\%$ reduction relative to \emph{Transformer-Direct}. Since \emph{PRISM} and \emph{Transformer-Direct} use the same encoder architecture, this additional gain cannot be attributed to the Transformer representation alone and instead supports the benefit of the probabilistic candidate generation and DT-guided refinement proposed in Section~\ref{sec:proposed_framework}. Similar gains are observed across all considered AP configurations.

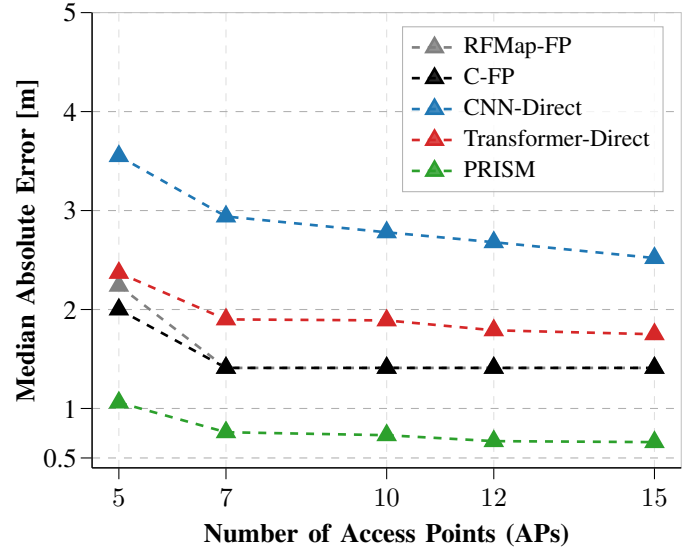
\begin{figure}[t]
\centering
\makebox[\columnwidth][c]{%
\begin{tikzpicture}

\definecolor{crimson2143940}{RGB}{214,39,40}
\definecolor{darkgray176}{RGB}{176,176,176}
\definecolor{forestgreen4416044}{RGB}{44,160,44}
\definecolor{lightgray204}{RGB}{204,204,204}
\definecolor{steelblue31119180}{RGB}{31,119,180}

\begin{axis}[
grid style={dashed, gray!30},
width=0.88\columnwidth,
height=0.68\columnwidth,
scale only axis,
xlabel style={font=\bfseries, yshift=2pt},
ylabel style={
    font=\bfseries,
    at={(axis description cs:-0.075,0.5)},
    anchor=south
},
axis lines=box,
axis line style={line width=0.25pt, draw=black!25},
extra description/.code={
    \draw[black, line width=0.55pt]
        (rel axis cs:0,0) -- (rel axis cs:1,0);
    \draw[black, line width=0.55pt]
        (rel axis cs:0,0) -- (rel axis cs:0,1);
},
tick align=outside,
tick pos=left,
tick style={line width=0.4pt, draw=black},
xtick style={color=black},
ytick style={color=black},
yticklabel style={xshift=2pt},
xmajorgrids,
ymajorgrids,
x grid style={gray!25},
y grid style={gray!25},
legend cell align={left},
legend style={
    at={(0.97,0.97)},
    anchor=north east,
    font=\small,
    draw=black!30,
    fill=white,
    fill opacity=0.85,
    text opacity=1
},
xlabel={\textbf{Number of Access Points (APs)}},
xmajorgrids,
xmin=4.5, xmax=15.5,
xtick={5,7,10,12,15},
xtick style={color=black},
y grid style={darkgray176},
ylabel={\textbf{Median Absolute Error [m]}},
ylabel style={
    at={(axis description cs:-0.075,0.5)},
    anchor=south
},
ymajorgrids,
ymin=0.4, ymax=5,
ytick={0.5,1,2,3,4,5},
yticklabels={0.5,1,2,3,4,5},
ytick style={color=black}
]
\addplot [
    line width=1.0pt,
    gray,
    dashed,
    mark=triangle*,
    mark size=3.5,
    mark options={solid}
]
coordinates {
    (5,2.24)
    (7,1.41)
    (10,1.41)
    (12,1.41)
    (15,1.41)
};
\addlegendentry{RFMap-FP}

\addplot [
    line width=1.0pt,
    black,
    dashed,
    mark=triangle*,
    mark size=3.5,
    mark options={solid}
]
coordinates {
    (5,2.00)
    (7,1.41)
    (10,1.41)
    (12,1.41)
    (15,1.41)
};
\addlegendentry{C-FP}
\addplot [
    line width=1.0pt,
    steelblue31119180,
    dashed,
    mark=triangle*,
    mark size=3.5,
    mark options={solid}
]
coordinates {
    (5,3.55)
    (7,2.94)
    (10,2.78)
    (12,2.68)
    (15,2.52)
};
\addlegendentry{CNN-Direct}

\addplot [
    line width=1.0pt,
    crimson2143940,
    dashed,
    mark=triangle*,
    mark size=3.5,
    mark options={solid}
]
coordinates {
    (5,2.37)
    (7,1.90)
    (10,1.89)
    (12,1.79)
    (15,1.75)
};
\addlegendentry{Transformer-Direct}

\addplot [
    line width=1.0pt,
    forestgreen4416044,
    dashed,
    mark=triangle*,
    mark size=3.5,
    mark options={solid}
]
coordinates {
    (5,1.06)
    (7,0.76)
    (10,0.73)
    (12,0.67)
    (15,0.66)
};
\addlegendentry{PRISM}

\end{axis}
\end{tikzpicture}%
}
\caption{Median absolute error versus number of APs for PRISM and benchmarks at an $\mathrm{SNR}_{\mathrm{obs}}$ of 25~dB.}
\label{fig:median_ap_PRISM_nn_fingerprinting}
\end{figure}

\begin{figure*}[!t]
\centering

\begin{minipage}{0.405\textwidth}
    \centering
    \includegraphics[width=\linewidth,trim={9cm 0cm 4cm 0cm},clip]
    {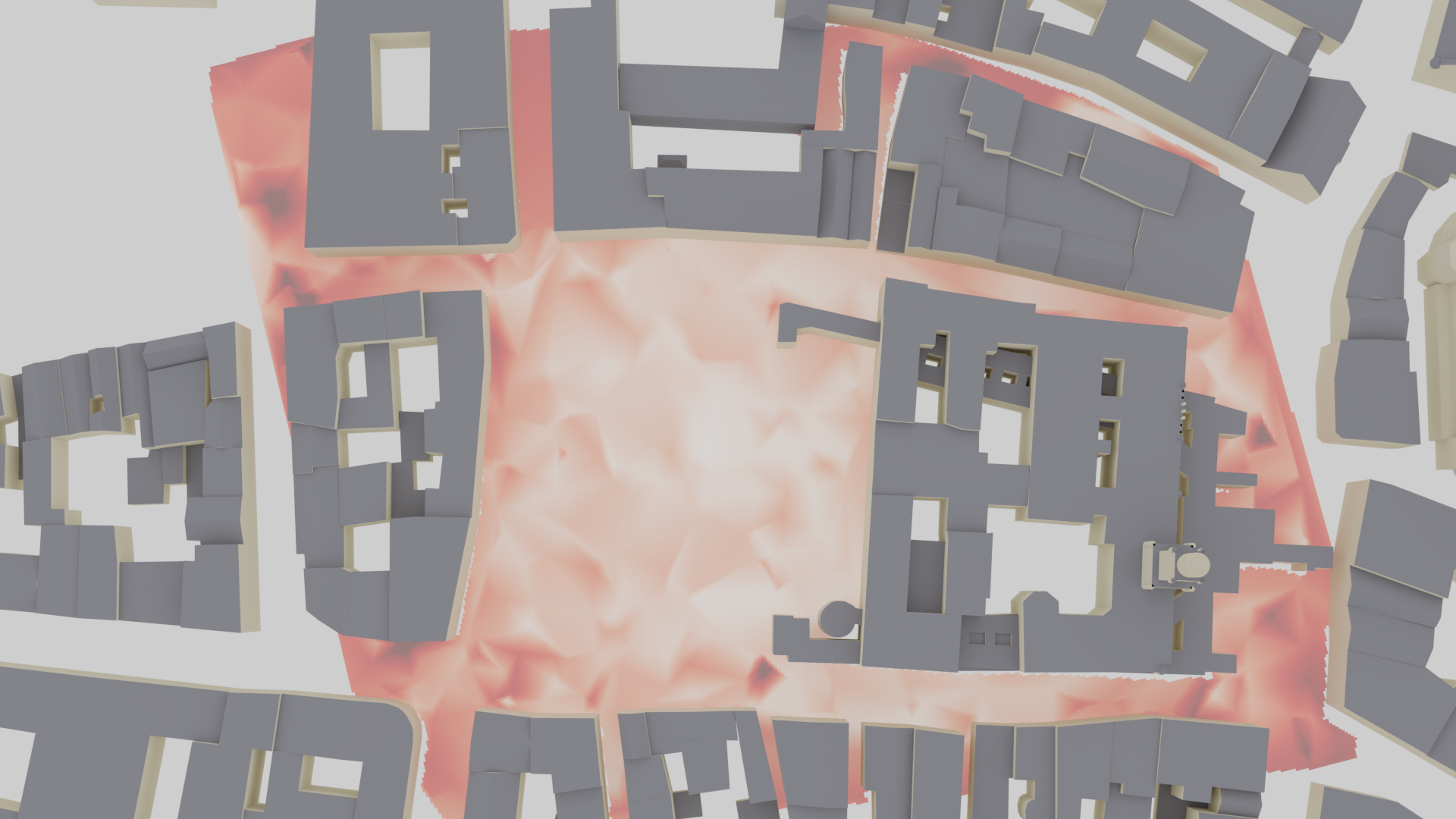}\\[-0.5ex]
    \footnotesize (a) \emph{Transformer-Direct}
\end{minipage}
\hspace{0.012\textwidth}
\begin{minipage}{0.405\textwidth}
    \centering
    \includegraphics[width=\linewidth,trim={9cm 0cm 4cm 0cm},clip]
    {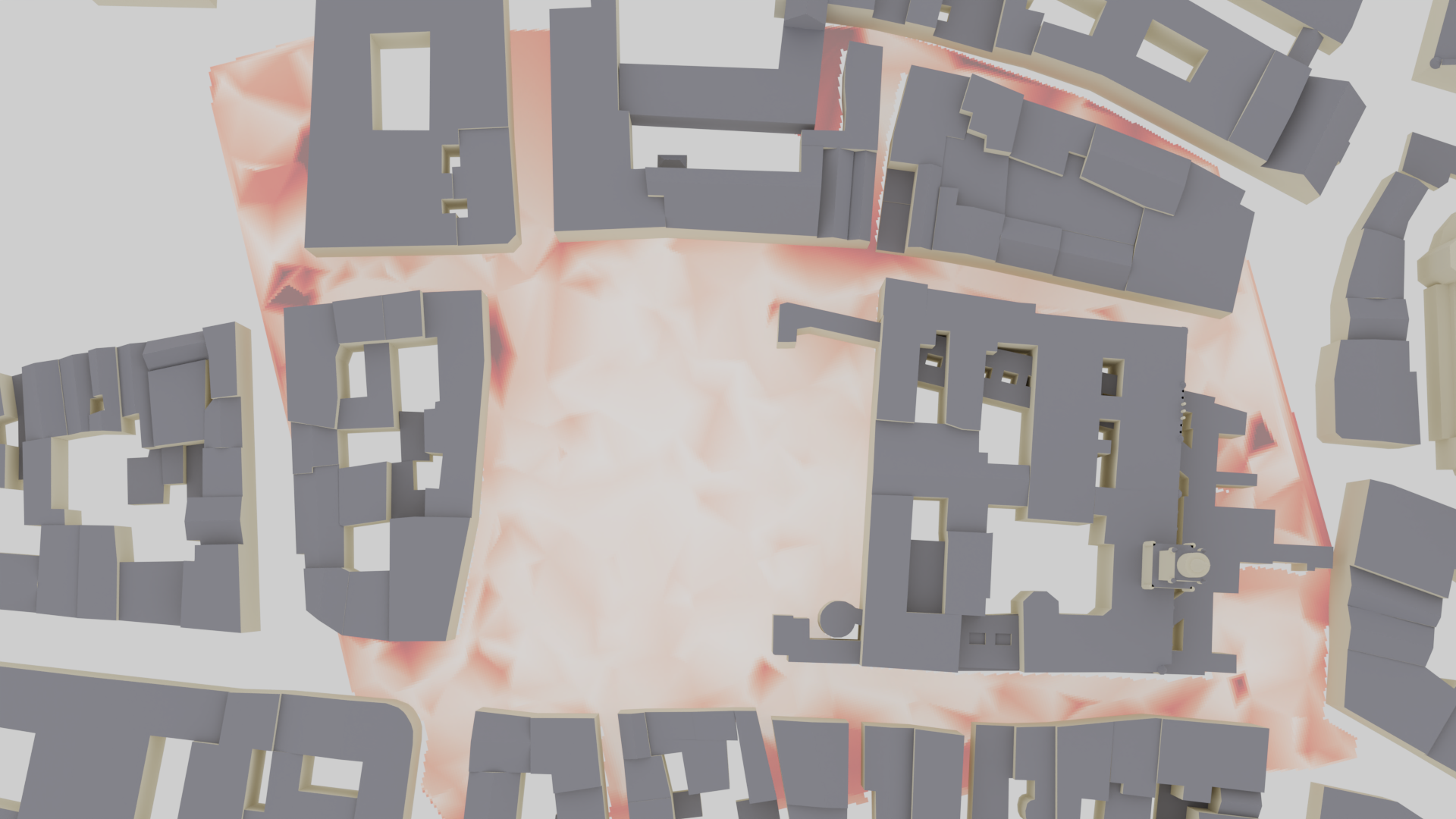}\\[-0.5ex]
    \footnotesize (b) \emph{PRISM}
\end{minipage}
\hspace{0.005\textwidth}
\begin{minipage}{0.055\textwidth}
    \centering
    \vspace{-0.2cm}
    \includegraphics[height=3.8cm,keepaspectratio]{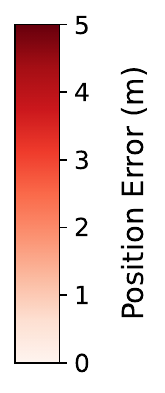}
\end{minipage}

\caption{Spatial distribution of localization error in the 10-AP Munich-scene configuration: (a) \emph{Transformer-Direct} and (b) \emph{PRISM}. The color scale indicates position error in meters.}
\label{fig:error_comparison}

\end{figure*}
We next examine whether increasing the fingerprint-database resolution can match the localization gains achieved by \emph{PRISM}. Fig.~\ref{fig:mae_ap_PRISM_fingerprinting_grids} shows that decreasing the fingerprint-grid spacing to $0.75\,\mathrm{m}$ and $0.5\,\mathrm{m}$ improves fingerprinting accuracy across the considered AP configurations. However, \emph{PRISM} remains more accurate than all fingerprinting schemes in every configuration, even when \emph{Dense-FP} uses a $0.5\,\mathrm{m}$ grid. At $10$ APs, \emph{Dense-FP} with $0.5\,\mathrm{m}$ spacing achieves an MAE of $2.17\,\mathrm{m}$, whereas \emph{PRISM} achieves $1.11\,\mathrm{m}$, corresponding to an approximately $49\%$ reduction. Increasing the fingerprint resolution improves accuracy at the cost of a larger database and exhaustive-search burden. Nevertheless, \emph{PRISM}
retains a clear accuracy advantage even over the $0.5\,\mathrm{m}$ \emph{Dense-FP}, supporting the benefit of combining learned
channel-to-location structure with DT-guided refinement beyond simply
increasing fingerprint-database resolution. In the remaining experiments, we use \emph{C-FP} with the default $1\,\mathrm{m}$ grid as the representative fingerprinting benchmark.

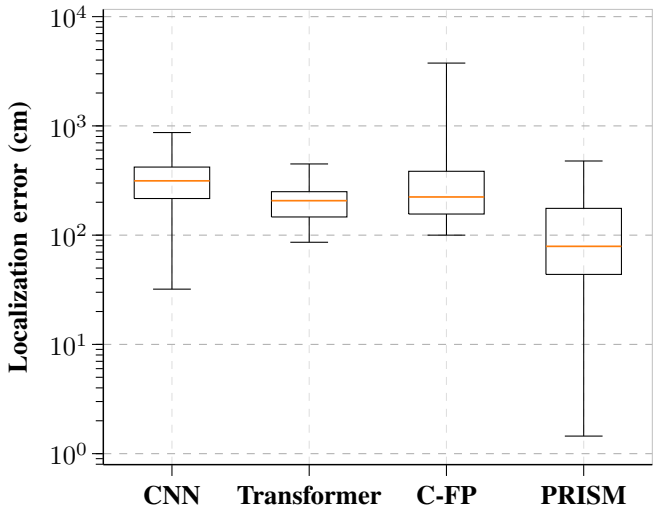
\begin{figure}

\begin{tikzpicture}

\definecolor{darkgray176}{RGB}{176,176,176}
\definecolor{darkorange25512714}{RGB}{255,127,14}

\begin{axis}[
grid style={dashed, gray!30},
width=0.82\columnwidth,
height=0.68\columnwidth,
scale only axis,
xlabel style={font=\bfseries, yshift=2pt},
ylabel style={
    font=\bfseries,
    at={(axis description cs:-0.075,0.5)},
    anchor=south
},
axis lines=box,
axis line style={line width=0.25pt, draw=black!25},
extra description/.code={
    \draw[black, line width=0.55pt]
        (rel axis cs:0,0) -- (rel axis cs:1,0);
    \draw[black, line width=0.55pt]
        (rel axis cs:0,0) -- (rel axis cs:0,1);
},
tick align=outside,
tick pos=left,
tick style={line width=0.4pt, draw=black},
xtick style={color=black},
ytick style={color=black},
yticklabel style={xshift=2pt},
xmajorgrids,
ymajorgrids,
x grid style={gray!25},
y grid style={gray!25},
legend cell align={left},
legend style={
    at={(0.97,0.97)},
    anchor=north east,
    font=\small,
    draw=black!30,
    fill=white,
    fill opacity=0.85,
    text opacity=1
},
xmin=0.5, xmax=4.5,
xtick style={color=black},
xtick={1,2,3,4},
xticklabels={\textbf{CNN},\textbf{Transformer},\textbf{C-FP},\textbf{PRISM}},
y grid style={darkgray176},
ylabel={\textbf{Localization error (cm)}},
ymin=0.793653693222124, ymax=11650.3921716102,
ymode=log,
ytick style={color=black}
]
\addplot [black]
coordinates {
(0.725,216.408153533936)
(1.275,216.408153533936)
(1.275,419.866912841797)
(0.725,419.866912841797)
(0.725,216.408153533936)
};
\addplot [black]
coordinates {
(1,216.408153533936)
(1,32.0440864562988)
};
\addplot [black]
coordinates {
(1,419.866912841797)
(1,869.030212402344)
};
\addplot [black]
coordinates {
(0.8625,32.0440864562988)
(1.1375,32.0440864562988)
};
\addplot [black]
coordinates {
(0.8625,869.030212402344)
(1.1375,869.030212402344)
};
\addplot [black]
coordinates {
(1.725,146.70906829834)
(2.275,146.70906829834)
(2.275,249.888832092285)
(1.725,249.888832092285)
(1.725,146.70906829834)
};
\addplot [black]
coordinates {
(2,146.70906829834)
(2,86.0808868408203)
};
\addplot [black]
coordinates {
(2,249.888832092285)
(2,448.466125488281)
};
\addplot [black]
coordinates {
(1.8625,86.0808868408203)
(2.1375,86.0808868408203)
};
\addplot [black]
coordinates {
(1.8625,448.466125488281)
(2.1375,448.466125488281)
};
\addplot [black]
coordinates {
(2.725,156.066017150879)
(3.275,156.066017150879)
(3.275,384.23291015625)
(2.725,384.23291015625)
(2.725,156.066017150879)
};
\addplot [black]
coordinates {
(3,156.066017150879)
(3,100)
};
\addplot [black]
coordinates {
(3,384.23291015625)
(3,3757.65869140625)
};
\addplot [black]
coordinates {
(2.8625,100)
(3.1375,100)
};
\addplot [black]
coordinates {
(2.8625,3757.65869140625)
(3.1375,3757.65869140625)
};
\addplot [black]
coordinates {
(3.725,43.7214860916138)
(4.275,43.7214860916138)
(4.275,175.865875244141)
(3.725,175.865875244141)
(3.725,43.7214860916138)
};
\addplot [black]
coordinates {
(4,43.7214860916138)
(4,1.44819498062134)
};
\addplot [black]
coordinates {
(4,175.865875244141)
(4,477.212585449219)
};
\addplot [black]
coordinates {
(3.8625,1.44819498062134)
(4.1375,1.44819498062134)
};
\addplot [black]
coordinates {
(3.8625,477.212585449219)
(4.1375,477.212585449219)
};
\addplot [line width=0.64pt, darkorange25512714]
coordinates {
(0.725,314.118362426758)
(1.275,314.118362426758)
};
\addplot [line width=0.64pt, darkorange25512714]
coordinates {
(1.725,206.768112182617)
(2.275,206.768112182617)
};
\addplot [line width=0.64pt, darkorange25512714]
coordinates {
(2.725,223.606796264648)
(3.275,223.606796264648)
};
\addplot [line width=0.64pt, darkorange25512714]
coordinates {
(3.725,79.0026092529297)
(4.275,79.0026092529297)
};
\end{axis}

\end{tikzpicture}
\caption{Localization error distribution for NLOS test points in the 10-AP Munich-scene configuration. The vertical axis is shown on a logarithmic scale.}
\label{fig:boxplots_nlos}
\end{figure}

Fig.~\ref{fig:median_ap_PRISM_nn_fingerprinting} shows that \emph{PRISM} also improves typical localization accuracy, achieving the lowest median error across all considered AP configurations. At $10$ APs, \emph{PRISM} reduces the median error from $1.90\,\mathrm{m}$ for \emph{Transformer-Direct} to approximately $0.70\,\mathrm{m}$, corresponding to a $63\%$ reduction. The median error of \emph{PRISM} further decreases to $0.66\,\mathrm{m}$ at $15$ APs. Comparing the mean and median errors shows that the fingerprinting benchmarks achieve competitive median accuracy but retain substantially larger mean errors, indicating a greater influence of high-error realizations. In contrast, \emph{PRISM} achieves the lowest values for both metrics, improving both average and typical localization accuracy.

To examine how the localization gains are distributed across the deployment area, Fig.~\ref{fig:error_comparison} compares the spatial errors of \emph{Transformer-Direct} and \emph{PRISM} for the $10$-AP configuration. \emph{Transformer-Direct} exhibits several extended high-error regions, particularly near scene boundaries and building edges. In contrast, \emph{PRISM} reduces both the intensity and spatial extent of these regions and provides more consistent accuracy across the deployment area. Although challenging regions remain for both schemes, the overall improvement indicates that DT-guided refinement helps resolve candidate locations that remain ambiguous under direct neural inference.

Focusing specifically on NLOS test locations, Fig.~\ref{fig:boxplots_nlos} compares the error distributions of \emph{PRISM} and the benchmark schemes. \emph{PRISM} achieves a median error of approximately $0.79\,\mathrm{m}$, compared with $2.07\,\mathrm{m}$ for \emph{Transformer-Direct}, $2.24\,\mathrm{m}$ for \emph{C-FP}, and $3.14\,\mathrm{m}$ for \emph{CNN-Direct}. Moreover, its $75$th-percentile error of approximately $1.76\,\mathrm{m}$ remains below the median error of every benchmark. Thus, the advantage of \emph{PRISM} persists across most NLOS test locations rather than being driven by a small subset of favorable realizations.

\subsection{Robustness to Observation SNR and DT Imperfections}
\label{subsec:robustness_noise_dt}

We evaluate the robustness of \emph{PRISM} to two practical mismatches: imperfect DT information during inference and a shift between the training and test observation SNRs. In all experiments, the trained models are reused without retraining under the modified test conditions.

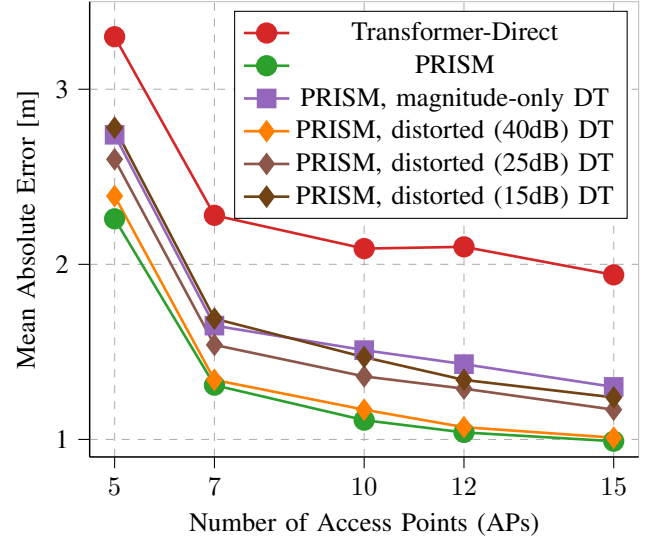
\begin{figure}[t]
\centering
\makebox[\columnwidth][c]{%
\begin{tikzpicture}

\definecolor{steelblue31119180}{RGB}{31,119,180}
\definecolor{crimson2143940}{RGB}{214,39,40}
\definecolor{darkgray176}{RGB}{176,176,176}
\definecolor{forestgreen4416044}{RGB}{44,160,44}
\definecolor{purple148103189}{RGB}{148,103,189}
\definecolor{brown1408675}{RGB}{140,86,75}

\begin{axis}[
grid style={dashed, gray!30},
width=0.82\columnwidth,
height=0.68\columnwidth,
scale only axis,
xlabel style={font=\bfseries, yshift=2pt},
ylabel style={
    font=\bfseries,
    at={(axis description cs:-0.075,0.5)},
    anchor=south
},
axis lines=box,
axis line style={line width=0.25pt, draw=black!25},
extra description/.code={
    \draw[black, line width=0.55pt]
        (rel axis cs:0,0) -- (rel axis cs:1,0);
    \draw[black, line width=0.55pt]
        (rel axis cs:0,0) -- (rel axis cs:0,1);
},
tick align=outside,
tick pos=left,
tick style={line width=0.4pt},
axis line style={line width=0.4pt},
x grid style={darkgray176},
xlabel={Number of Access Points (APs)},
xmajorgrids,
xmin=4.5, xmax=15.5,
xtick={5,7,10,12,15},
xtick style={color=black},
y grid style={darkgray176},
ylabel={Mean Absolute Error [m]},
ylabel style={
    at={(axis description cs:-0.075,0.5)},
    anchor=south
},
ymajorgrids,
ymin=0.9, ymax=3.5,
ytick={1,2,3,4},
yticklabels={1,2,3,4},
ytick style={color=black}
]

\addplot [
    line width=1.0pt,
    crimson2143940,
    mark=*,
    mark size=3.5,
    mark options={solid}
]
coordinates {
    (5,3.30)
    (7,2.28)
    (10,2.09)
    (12,2.10)
    (15,1.94)
};
\addlegendentry{Transformer-Direct}

\addplot [
    line width=1.0pt,
    forestgreen4416044,
    mark=*,
    mark size=3.5,
    mark options={solid}
]
coordinates {
    (5,2.26)
    (7,1.31)
    (10,1.11)
    (12,1.04)
    (15,0.99)
};
\addlegendentry{PRISM}

\addplot [
    line width=1.0pt,
    purple148103189,
    mark=square*,
    mark size=3.2,
    mark options={solid}
]
coordinates {
    (5,2.74)
    (7,1.65)
    (10,1.51)
    (12,1.43)
    (15,1.30)
};
\addlegendentry{PRISM, magnitude-only DT}

\addplot [
    line width=1.0pt,
    orange,
    mark=diamond*,
    mark size=3.5,
    mark options={solid}
]
coordinates {
    (5,2.39)
    (7,1.34)
    (10,1.17)
    (12,1.07)
    (15,1.01)
};
\addlegendentry{PRISM, distorted (40dB) DT}

\addplot [
    line width=1.0pt,
    brown1408675,
    mark=diamond*,
    mark size=3.5,
    mark options={solid}
]
coordinates {
    (5,2.60)
    (7,1.54)
    (10,1.36)
    (12,1.29)
    (15,1.17)
};
\addlegendentry{PRISM, distorted (25dB) DT}
\addplot [
    line width=1.0pt,
    sepia,
    mark=diamond*,
    mark size=3.5,
    mark options={solid}
]
coordinates {
    (5,2.78)
    (7,1.69)
    (10,1.47)
    (12,1.34)
    (15,1.24)
};
\addlegendentry{PRISM, distorted (15dB) DT}

\end{axis}
\end{tikzpicture}%
}
\caption{Mean absolute error versus number of APs under imperfect DT information. The observed CIR features are generated at $\mathrm{SNR}_{\mathrm{obs}}=25$~dB.}
\label{fig:mae_ap_dt_imperfections}
\end{figure}

\subsubsection{Robustness to Imperfect DT Information}

We first evaluate how accurately the DT must reproduce the channel
features for \emph{PRISM} to remain effective. Fig.~\ref{fig:mae_ap_dt_imperfections} compares the standard configuration
with a magnitude-only DT and DT features distorted at $\mathrm{SDR}_{\mathrm{DT}}\in\{15,25,40\}$~dB. The observed CIR features
are generated at $\mathrm{SNR}_{\mathrm{obs}}=25$~dB in all cases.

Fig.~\ref{fig:mae_ap_dt_imperfections} shows graceful degradation as
the DT information becomes less accurate. At $10$ APs, the MAE increases
from $1.11\,\mathrm{m}$ with the full DT to $1.17\,\mathrm{m}$,
$1.36\,\mathrm{m}$, and $1.47\,\mathrm{m}$ for
$\mathrm{SDR}_{\mathrm{DT}}=40$, $25$, and $15$~dB, respectively, and to
$1.51\,\mathrm{m}$ with magnitude-only DT features. Nevertheless, every
imperfect-DT configuration remains more accurate than
\emph{Transformer-Direct} across all considered AP configurations. At
$10$ APs, even the magnitude-only case improves on its
$2.09\,\mathrm{m}$ MAE by approximately $28\%$. More generally, the results suggest that \emph{PRISM} does not require exact channel reproduction, provided that the imperfect DT preserves useful relative consistency among candidate locations.

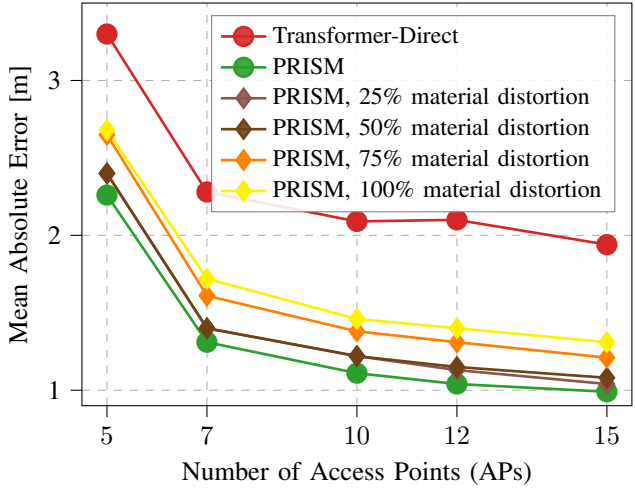
\begin{figure}[t]
\centering
\makebox[\columnwidth][c]{%
\begin{tikzpicture}

\definecolor{steelblue31119180}{RGB}{31,119,180}
\definecolor{crimson2143940}{RGB}{214,39,40}
\definecolor{darkgray176}{RGB}{176,176,176}
\definecolor{forestgreen4416044}{RGB}{44,160,44}
\definecolor{purple148103189}{RGB}{148,103,189}
\definecolor{brown1408675}{RGB}{140,86,75}

\begin{axis}[
grid style={dashed, gray!30},
width=1.00\columnwidth,
height=0.78\columnwidth,
legend cell align={left},
legend style={
    at={(0.97,0.97)},
    anchor=north east,
    font=\small,
    draw=black!50,
    fill=white,
    fill opacity=0.9,
    text opacity=1
},
tick align=outside,
tick pos=left,
tick style={line width=0.4pt},
axis line style={line width=0.4pt},
x grid style={darkgray176},
xlabel={Number of Access Points (APs)},
xmajorgrids,
xmin=4.5, xmax=15.5,
xtick={5,7,10,12,15},
xtick style={color=black},
y grid style={darkgray176},
ylabel={Mean Absolute Error [m]},
ylabel style={
    at={(axis description cs:-0.075,0.5)},
    anchor=south
},
ymajorgrids,
ymin=0.9, ymax=3.5,
ytick={1,2,3,4},
yticklabels={1,2,3,4},
ytick style={color=black}
]

\addplot [
    line width=1.0pt,
    crimson2143940,
    mark=*,
    mark size=3.5,
    mark options={solid}
]
coordinates {
    (5,3.30)
    (7,2.28)
    (10,2.09)
    (12,2.10)
    (15,1.94)
};
\addlegendentry{Transformer-Direct}

\addplot [
    line width=1.0pt,
    forestgreen4416044,
    mark=*,
    mark size=3.5,
    mark options={solid}
]
coordinates {
    (5,2.26)
    (7,1.31)
    (10,1.11)
    (12,1.04)
    (15,0.99)
};
\addlegendentry{PRISM}

\addplot [
    line width=1.0pt,
    brown1408675,
    mark=diamond*,
    mark size=3.5,
    mark options={solid}
]
coordinates {
    (5,2.4)
    (7,1.4)
    (10,1.22)
    (12,1.13)
    (15,1.04)
};
\addlegendentry{PRISM, 25\% material distortion}
\addplot [
    line width=1.0pt,
    sepia,
    mark=diamond*,
    mark size=3.5,
    mark options={solid}
]
coordinates {
    (5,2.4)
    (7,1.4)
    (10,1.22)
    (12,1.15)
    (15,1.08)
};
\addlegendentry{PRISM, 50\% material distortion}
\addplot [
    line width=1.0pt,
    orange,
    mark=diamond*,
    mark size=3.5,
    mark options={solid}
]
coordinates {
    (5,2.65)
    (7,1.61)
    (10,1.38)
    (12,1.31)
    (15,1.21)
};
\addlegendentry{PRISM, 75\% material distortion}
\addplot [
    line width=1.0pt,
    yellow,
    mark=diamond*,
    mark size=3.5,
    mark options={solid}
]
coordinates {
    (5,2.68)
    (7,1.72)
    (10,1.46)
    (12,1.4)
    (15,1.31)
};
\addlegendentry{PRISM, 100\% material distortion}

\end{axis}
\end{tikzpicture}%
}
\caption{Mean absolute error versus number of APs under imperfect DT information. The observed CIR features are generated at $\mathrm{SNR}_{\mathrm{obs}}=25$~dB. For each distorted-DT case, the material properties of a randomly
selected fraction of scene objects are replaced by those of $20\,\mathrm{cm}$ brick.}
\label{fig:mae_ap_dt_mat_imperfections}
\end{figure}

We further evaluate a structured DT mismatch in which the material
properties of a randomly selected fraction of scene objects are replaced by
those of $20\,\mathrm{cm}$ brick. Fig.~\ref{fig:mae_ap_dt_mat_imperfections}
shows the resulting MAE as the replacement fraction increases.

Fig.~\ref{fig:mae_ap_dt_mat_imperfections} shows graceful degradation
as the material mismatch increases. At $15$ APs, the MAE increases from
$0.99\,\mathrm{m}$ with the full DT to $1.04\,\mathrm{m}$,
$1.08\,\mathrm{m}$, $1.21\,\mathrm{m}$, and $1.31\,\mathrm{m}$ for
$25\%$, $50\%$, $75\%$, and $100\%$ material replacement, respectively.
Nevertheless, even the $100\%$ replacement case remains more accurate than \emph{Transformer-Direct} across every considered AP configuration. This indicates that the preserved scene geometry and remaining relative channel-consistency structure can still provide effective DT guidance under
severe material-property mismatch.

Auxiliary experiments with large movable objects, such as trucks, indicated that the resulting DT mismatch was mainly localized and produced no noticeable change in the aggregate mean or median localization error. No clear degradation was observed even for nearby test locations in the considered setting. A systematic evaluation under more dynamic scene changes is left for future work.

\subsubsection{Robustness to Observation SNR Mismatch}

We now evaluate whether the localization schemes remain robust when the test observation SNR differs from the setting used during training. Using the $10$-AP configuration, models trained at $15$~dB are tested at $15$~dB and $25$~dB, while models trained at $25$~dB are tested at $25$~dB and $40$~dB. For \emph{C-FP}, the first SNR in each pair is applied to the fingerprint database and the second to the test observations.

\begin{table}[htbp]
\centering
\footnotesize
\renewcommand{\arraystretch}{1.5}
\caption{MAE under observation-SNR mismatch in the 10-AP configuration.}
\label{tab:noise_comparison}
\begin{tabular}{l cc cc}
\toprule
& \multicolumn{2}{c}{%
    \shortstack{\textbf{Training/database}\\\textbf{SNR: 15\,dB}}}
& \multicolumn{2}{c}{%
    \shortstack{\textbf{Training/database}\\\textbf{SNR: 25\,dB}}} \\
\cmidrule(lr){2-3}\cmidrule(lr){4-5}
\textbf{Algorithm}
& \textbf{15\,dB}
& \textbf{25\,dB}
& \textbf{25\,dB}
& \textbf{40\,dB} \\
\midrule
CNN-Direct         & 3.56 & 12.56 & 3.30 & 8.68 \\
Transformer-Direct & 1.85 & 5.93  & 2.09 & 2.76 \\
C-FP               & 3.66 & 5.17  & 3.71 & 4.75 \\
PRISM              & \textbf{1.25} & \textbf{1.94 } & \textbf{1.11} & \textbf{1.59} \\
\bottomrule
\end{tabular}
\end{table}

\begin{figure*}[t]
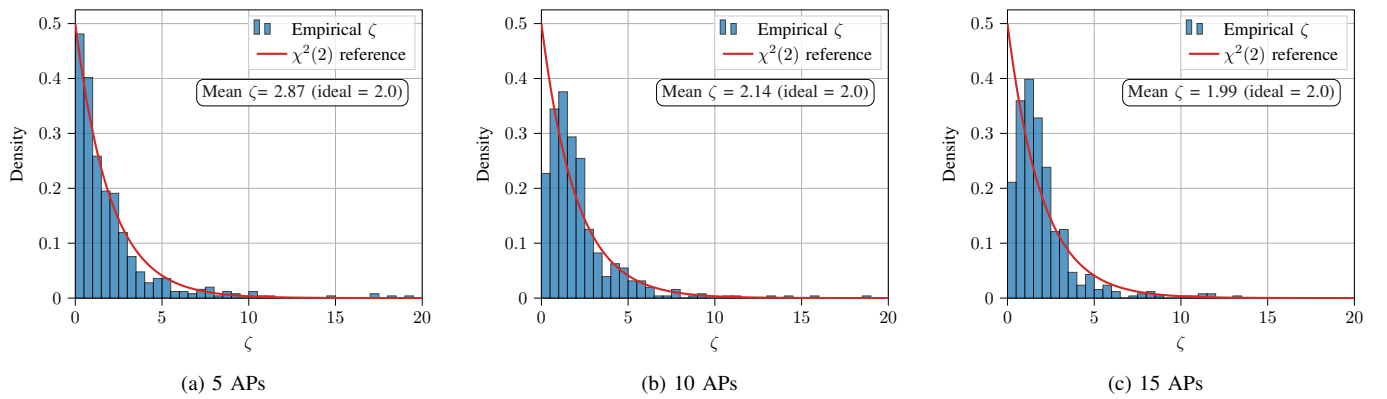

    \centering
    \begin{subfigure}[b]{0.32\textwidth}
        \centering
        \resizebox{\linewidth}{!}{\input{Figures/chi2_5AP.tex}}
        \caption{5 APs}
        \label{fig:chi2-5ap}
    \end{subfigure}
    \hfill
    \begin{subfigure}[b]{0.32\textwidth}
        \centering
        \resizebox{\linewidth}{!}{\input{Figures/chi2_10AP.tex}}
        \caption{10 APs}
        \label{fig:chi2-10ap}
    \end{subfigure}
    \hfill
    \begin{subfigure}[b]{0.32\textwidth}
        \centering
        \resizebox{\linewidth}{!}{\input{Figures/chi2_15AP.tex}}
        \caption{15 APs}
        \label{fig:chi2-15ap}
    \end{subfigure}
    \caption{
    Empirical distribution of the squared Mahalanobis statistic $\zeta$ compared with the $\chi^2(2)$ reference distribution for different numbers of APs. The red curve shows the ideal $\chi^2(2)$ distribution,
    while the blue bars show the observed empirical distribution of the
    squared Mahalanobis statistic.}
    \label{fig:chi2-comparison}
\end{figure*}
Table~\ref{tab:noise_comparison} shows that increasing the test SNR
does not necessarily improve localization when it differs from the SNR used for training or fingerprint-database construction. Although the observations contain less noise, the change in SNR shifts their feature distribution from that encountered during training. This effect is most pronounced for \emph{CNN-Direct}, whose MAE increases from $3.56\,\mathrm{m}$ to $12.56\,\mathrm{m}$ under the $15$-to-$25$~dB mismatch and from $3.30\,\mathrm{m}$ to $8.68\,\mathrm{m}$ under the
$25$-to-$40$~dB mismatch. \emph{Transformer-Direct} is more stable,
particularly in the latter case, while \emph{C-FP} exhibits more moderate
degradation but retains a comparatively high MAE.

In contrast, \emph{PRISM} maintains an MAE below $2\,\mathrm{m}$ in
all tested conditions and exhibits the smallest absolute degradation in both mismatch cases. Its MAE increases from $1.25\,\mathrm{m}$ to
$1.94\,\mathrm{m}$ under the $15$-to-$25$~dB mismatch and from
$1.11\,\mathrm{m}$ to $1.59\,\mathrm{m}$ under the $25$-to-$40$~dB mismatch. These results suggest that DT-guided candidate scoring provides an additional site-specific consistency check that reduces the sensitivity of the final localization estimate to changes in the observation statistics.

\subsection{Uncertainty Quantification}
\label{subsec:uncertainty_quantification}

In addition to a point estimate, \emph{PRISM} outputs a predictive covariance that quantifies the uncertainty associated with its learned location proposal. We evaluate whether this covariance is consistent with
the observed localization errors using the squared Mahalanobis statistic. To assess calibration, we compute the squared Mahalanobis statistic
\begin{align*}
\zeta^{(m)}
=
\left(\boldsymbol u^{(m)}-\boldsymbol\mu^{(m)}\right)^{\mathsf T}
\left(\boldsymbol\Sigma^{(m)}\right)^{-1}
\left(\boldsymbol u^{(m)}-\boldsymbol\mu^{(m)}\right),
\end{align*} where $\boldsymbol u^{(m)}$ is the ground-truth UE location and
$\boldsymbol\mu^{(m)}$ and $\boldsymbol\Sigma^{(m)}$ are the predicted mean and covariance for test sample $m$. If the two-dimensional Gaussian proposal is calibrated, $\zeta^{(m)}$ approximately follows a $\chi^2$ distribution with two degrees of freedom. Fig.~\ref{fig:chi2-comparison} compares its empirical distribution with the $\chi^2(2)$ reference for different numbers of APs.

Fig.~\ref{fig:chi2-comparison} shows that the empirical distribution of $\zeta$ approaches the $\chi^2(2)$ reference as the number of APs increases. Its mean decreases from $2.87$ for $5$ APs to $2.14$ for $10$ APs and $1.99$ for $15$ APs, compared with the reference mean of $2$. The $5$-AP case is overconfident on average and exhibits a heavier upper tail, whereas the $10$-AP distribution is closer to the reference and the $15$-AP case is well calibrated in the mean. These results indicate that the predicted covariance becomes increasingly consistent with the observed localization errors as more AP observations are available.

Table~\ref{tab:uncertainty_calibration} provides a complementary
coverage-based assessment for the $15$-AP configuration. The mean statistic
of $1.992$ is close to the $\chi^2(2)$ reference mean of $2$, while the
empirical coverages of $93.36\%$, $96.88\%$, and $98.63\%$ are close to or
slightly above the nominal  $90\%$, $95\%$, and $99\%$ levels, respectively. Thus, the predicted covariance is well calibrated in
the mean and provides near-nominal coverage at moderate-to-high confidence
levels, although some deviation remains across the full distribution.

\begin{table}[htbp]
    \centering
    \caption{Empirical proposal uncertainty coverage for PRISM in the 15-AP
    configuration using the squared Mahalanobis statistic.}
    \label{tab:uncertainty_calibration}
    \begin{tabular}{lrr}
        \toprule
        \textbf{Confidence Level}
            & \textbf{Expected Coverage}
            & \textbf{Actual Coverage} \\
            & (\%) & (\%) \\
        \midrule
        50\%  & 50.0  & 42.58 \\
        68\%  & 68.0  & 71.48 \\
        90\%  & 90.0  & 93.36 \\
        95\%  & 95.0  & 96.88 \\
        99\%  & 99.0  & 98.63 \\
        \bottomrule
    \end{tabular}

    \vspace{0.8em}
    \small
    \textbf{Squared Mahalanobis statistics:}
    Mean = 1.992,\quad
    Median = 1.54,\quad
    90th percentile = 3.689.
\end{table}

\section{Conclusion}
\label{section:conclusion}

This paper proposed a DT-guided probabilistic inverse learning framework for site-specific wireless localization using distributed single-antenna APs and a site-specific digital twin (DT). The localization task was formulated as a
probabilistic inverse problem, where a Transformer-based model learns a
conditional proposal distribution over UE locations from multi-node CIR
observations. Instead of treating localization as a purely black-box mapping, the proposed framework combines probabilistic inverse learning with DT-based forward-consistency refinement. The proposal spread is regularized during training to control finite candidate generation, while at inference the DT explicitly reweights sampled candidates according to their channel consistency.

Numerical experiments in realistic site-specific ray-traced environments demonstrated that the proposed framework consistently improves localization
accuracy over classical fingerprinting and purely data-driven neural
benchmarks. In particular, the proposed scheme achieved up to $63\%$ lower
median localization error relative to a Transformer-based benchmark and
about $0.7\,\mathrm{m}$ median error using only 10 distributed
single-antenna APs. The results also showed robustness to observation noise
and non-ideal DT conditions, including magnitude-only DT information.
Furthermore, the probabilistic formulation provided useful predictive
uncertainty estimates, with empirical calibration improving as the number of
APs increased.

Future work includes extending the framework from distributed single-antenna APs to multi-antenna sensing and three-dimensional localization, including UAV and industrial sensing scenarios, and studying online adaptation of the DT-assisted localization scheme in dynamic environments.  Another promising direction is to combine the proposed
DT-guided probabilistic inference framework with federated or distributed data
collection, enabling scalable site-specific localization with reduced
measurement overhead and improved adaptability to changing propagation
conditions.

\bibliographystyle{IEEEtran}
\bibliography{references}

\end{document}